\documentclass[12pt]{article}
\pdfoutput=1
\usepackage[a4paper]{geometry}
 
\usepackage[T1]{fontenc} 
\usepackage{comment}

\usepackage{jheppub,hyperref,float,array,adjustbox,mathtools,physics, xcolor}
\usepackage{lipsum}
\usepackage{booktabs}
\usepackage{pdflscape}
\usepackage{geometry}
\usepackage{fancyhdr}
\usepackage{changepage}
\usepackage[english]{babel}

\usepackage{xcolor,tikz,pgfplots,amsmath,amsfonts}
\usepackage{graphicx}
\usepackage[mathscr]{euscript}
\usetikzlibrary{matrix,calc,positioning,decorations.markings,decorations.pathmorphing,decorations.pathreplacing}
\usetikzlibrary{arrows,cd}
\usetikzlibrary{shapes.misc}
\usetikzlibrary {shapes.geometric}
\usetikzlibrary {shapes.multipart}
\usetikzlibrary{decorations.markings}
\usetikzlibrary{backgrounds}
\usetikzlibrary{shapes.gates.logic.US}
\usetikzlibrary{circuits.ee.IEC}
\usepackage{bbding}

\definecolor{terradisiena}{RGB}{233,116,81}
\definecolor{strisciadipietro}{RGB}{229,204,255}
\definecolor{verdepetrolio}{RGB}{33,100,119}
\definecolor{SUcol}{rgb}{0.5, 0.85, 0.8} 

\usepackage{tikz}
\usetikzlibrary{automata,positioning,calc}
\usetikzlibrary{decorations.markings}
\tikzset{->-/.style={decoration={markings, mark=at position #1 with {\arrow{>}}},postaction={decorate}}}
\tikzset{-<-/.style={decoration={markings, mark=at position #1 with {\arrow{<}}},postaction={decorate}}}
\tikzset{auto shift/.style={auto=right,->, to path={ let \p1=(\tikztostart),\p2=(\tikztotarget), \n1={atan2(\y2-\y1,\x2-\x1)},\n2={\n1+180} in ($(\tikztostart.{\n1})!1mm!270:(\tikztotarget.{\n2})$) -- ($(\tikztotarget.{\n2})!1mm!90:(\tikztostart.{\n1})$) \tikztonodes}}}

\usetikzlibrary {shapes.geometric}

\DeclareMathOperator{\SU}{SU}
\DeclareMathOperator{\UU}{U}

\newcommand{\numberset}{\mathbb}

\newcommand{\Z}{\numberset{Z}}

\newcommand{\C}{\numberset{C}}

\newcommand{\ee}{\mathrm{e}}
\newcommand{\mi}{\mathrm{i}}
\renewcommand{\Re}{\operatorname{Re}}
\renewcommand{\Im}{\operatorname{Im}}

\newcolumntype{?}{!{\vrule width 0.4pt}}

\usepackage{float}

\title{
\begin{center}
Exact results on the Bethe Ansatz evaluation of the SCI
\end{center}
}
	
\author[a]{Antonio Amariti,}
\author[a,b]{Pietro Glorioso,}

\affiliation[a]{INFN, Sezione di Milano, Via Celoria 16, I-20133 Milano, Italy}
\affiliation[b]{Dipartimento di Fisica, Università degli studi di Milano, Via Celoria 16, I-20133, Milano, Italy}

\emailAdd{antonio.amariti@mi.infn.it}
\emailAdd{pietro.glorioso@mi.infn.it}

\abstract{
We evaluate the superconformal index using the Bethe Ansatz (BA) approach for  4d $\mathcal{N}=1$ toric  quiver gauge theories with a small amount of gauge groups.
We restrict to $\mathrm{SU}(2)$ gauge factors and compare the results with the ones obtained by a direct evaluation of the index.
The answer obtained from the BA approach using only discrete solutions for the BA equations does not always reproduce the direct evaluation result.
A similar problem affects the case of $\mathrm{SU}(N)$ $\mathcal{N}=4$ SYM for $N \geq 3$, and it requires to introduce continuous solutions to the BA equations.
However, we find  that for $T^{1,1}$ and partially for the suspended pinch point  the matching is exact in absence of the contributions from continuous solutions.  
}

\begin{document}
\maketitle
\flushbottom
\allowdisplaybreaks

\section{Introduction}


The superconformal index (SCI) \cite{Kinney:2005ej,Romelsberger:2005eg} of 4d $\mathcal{N}=1$ supersymmetric field theories is a useful quantity that in the last two decades allowed to access to exact non-perturbative results in the analysis of SCFTs.
For example, it played a prominent role in the microstate counting program of 5d rotating black holes as shown in \cite{Benini:2018ywd,Choi:2018hmj}.
In general the evaluation of the SCI is a non-trivial task,  despite the fact that the functional integral is reduced to a matrix integral.
Nevertheless, it is still possible to reformulate the problem in terms of an exact expansion using the formalism spelled out in \cite{Closset:2017bse}.
Such expansion, proposed in \cite{Benini:2018mlo} consists of evaluating the integral as an expansion over the solutions of a set of transcendental equations named Bethe Ansatz Equations (BAEs). 
Along this line, in \cite{Benini:2018ywd}, the  SCI for $\SU(N)$ $\mathcal{N}=4$ SYM was then evaluated at large $N$ and complex fugacities, matching and generalizing the results obtained from other approaches \cite{Honda:2019cio,ArabiArdehali:2019orz,ArabiArdehali:2019tdm,Cabo-Bizet:2019osg,Amariti:2019mgp,Kim:2019yrz,Amariti:2020jyx,Benini:2020gjh,Choi:2023tiq,Cabo-Bizet:2020nkr,Cabo-Bizet:2018ehj,Amariti:2023rci,Jejjala:2021hlt,Ardehali:2021irq,Goldstein:2020yvj,Cabo-Bizet:2019eaf,Choi:2022asl,ArabiArdehali:2021nsx,Cassani:2021fyv,Amariti:2021ubd}, and reproducing the expected results from the counting of the microstates of the dual black hole.

These results were then generalized, for example by comparing the various saddles obtained from the gravitational expansion \cite{Aharony:2021zkr,Mamroud:2022msu,Aharony:2024ntg} or by computing the expected entropy for models with a lower amount of supersymmetry \cite{Lezcano:2019pae,Lanir:2019abx,GonzalezLezcano:2020yeb,Benini:2020gjh}. 
Recently, in order to probe the Hawking page transition (see also \cite{Copetti:2020dil} for interesting results in this direction) the index has been evaluated in presence of Gukov-Witten defects as well, and the results matched with the ones expected from holography \cite{Chen:2023lzq,Cabo-Bizet:2023ejm,Amariti:2024bsr}.

Despite such successes the evaluation of the index through this formalism at small rank is more challenging \cite{Agarwal:2020zwm,Murthy:2020scj,Benini:2021ano,Lezcano:2021qbj,Deddo:2025jrg,Choi:2025lck}.
For example, the index was evaluated using the BA approach in \cite{Benini:2021ano,Lezcano:2021qbj} for $\SU(2)$ and $\SU(3)$ $\mathcal{N}=4$ SYM. In such cases the matching with the expected result (at small fugacities), obtained from a
direct evaluation, has been found only for the $\SU(2)$ case, while, already for $N=3$ it has been observed that achieving this agreement requires more work.
The basic reason behind such observation is that the \emph{discrete} solutions to the BAEs are not enough to guarantee a complete answer, and further 
\emph{continuous} solutions must necessarily taken into account. 

Recently a general formalism has been proposed to solve such a challenge in \cite{Cabo-Bizet:2024kfe}, and, in this way, it has been shown that the expansion obtained for $\SU(3)$  
$\mathcal{N}=4$ SYM is consistent with the results expected from the direct evaluation.
Namely it has been shown that the divergent (a.k.a. tachyonic) \cite{Cabo-Bizet:2024kfe} contributions from the discrete Bethe roots cancel against analogous contributions due to the continuous solutions. Furthermore it has been shown also that the contribution of the continuous solutions to the index is localized, \emph{i.e.} the index does not pick up contributions from entire manifolds of holonomies, but only from discrete points on such manifolds.
Beyond the $\SU(N)$ $\mathcal{N}=4$ SYM case little is known for the small rank expansion of the SCI and it is not known, so far, if and when the exact evaluation of the index is possible in terms of discrete solutions or where the continuous solutions must be considered.

Motivated by this discussion in this paper we initiate a program to fill such a gap in the literature, by considering examples of increasing complexity.
The systematic analysis of the SCI at low rank gives us a spectroscopy that may be useful to shed more light on the BAEs expansion of 
the index at small ranks.
Here we restrict ourselves to toric quiver gauge theories, \emph{i.e.} quiver with products of $\SU(N)$ gauge factors that describe stack of D3 branes probing CY$_3$ singularities.
We consider only $\SU(2)$ gauge groups and we focus on the cases with $2$ and $3$ gauge groups.
Despite the simplicity of the setup, interesting features already emerge at this level, as we detail in the following.

First, we find that  the numerical expansion of the index over the discrete Bethe roots does not always reproduce the expected result.
This may be either due to the fact that other “exotic” discrete solutions contribute to the SCI or to the presence of continuous solutions.
While the first possibility does not seem quite plausible, due to the genericity of the ansatz used to scan over possible sets of solutions, the second one is compatible with the findings of \cite{Benini:2021ano,Cabo-Bizet:2024kfe}, where continuous solutions are expected if more than one holonomy is integrated over in the index.
However, more surprisingly in view of this last observation, we also found cases where the sum over the set of the discrete Bethe roots exactly reproduces the direct expansion of the SCI.

For example, we found that it is indeed the case for the conifold and, for non-generic flavor fugacities, also for the suspended pinch point (SPP) singularity.
In the following we provide a detailed analysis of these cases, by evaluating the index contribution of each discrete Bethe root and providing the matching with the expected results.
A common property of the evaluation over the various roots, as already observed in \cite{Benini:2021ano,Cabo-Bizet:2024kfe} for $\SU(N)$ $\mathcal{N}=4$ SYM, is that each one carries a tachyonic contribution, which cancels only after taking the sum over the various roots.

The first model where we found such matching corresponds to the conifold.
In this case we have solved the BAEs explicitly, finding the set of discrete inequivalent solutions.
Such set includes the Hong-Liu (HL) solutions \cite{Hong:2018viz} (see also \cite{Hosseini:2016cyf}), as well as additional new ones.
Then we have shown that the index evaluated using the BA approach on such solutions reproduces the result obtained from direct evaluation, for generic values of the flavor fugacities.
Each solution gives rise to divergent contributions, but such divergences cancel one each other once the final sum is considered with the correct amount of degenerations. 

The second model where the direct evaluation and the BA approach coincide corresponds to the SPP quiver gauge theory.
In this case there are three gauge groups and five abelian global symmetries. 
Actually, in this case we have not been able to solve the BAEs explicitly, but, inspired by the case of the conifold, we generated a set of solutions including the HL solutions and some additional elements along the lines of the results of $T^{1,1}$.
We have first verified that such solutions solve the system of BAEs and then we have evaluated the SCI over such discrete set.
In this case we have found that the divergent terms do not cancel  among each other for generic values of the flavor fugacities.
This is a signal that some discrete or continuous solutions is missing.
However, we have observed that there is a well defined limit associated to turning off some flavor fugacities, where such divergent terms cancel and where the index obtained from the BA approach coincides with the one obtained from direct evaluation. 

\section{The Bethe Ansatz approach}
\label{barew}

In this Section we briefly review the Bethe Ansatz (BA) approach.
This techinique was first introduced in \cite{Closset:2017bse}, and then generalized in \cite{Benini:2018mlo}, to rewrite and compute the SCI.
A comprehensive review can be found in \cite{Benini:2018mlo}.

The SCI of a 4d $\mathcal{N}=1$ gauge theory counts the local operators in short representations of the superconformal algebra $\mathfrak{su}(2,2|1)$.
Chosen a complex supercharge $\mathcal{Q}$, the SCI is defined as a refined Witten index of the theory on $S^3$ \cite{Kinney:2005ej,Romelsberger:2005eg}:
\begin{equation}
    \label{SCIdef}
    \mathcal{I}\equiv \Tr_{\mathcal{H}_{S^3}} (-1)^F 
    \ee^{-\beta \{\mathcal{Q},\mathcal{Q}^\dag\} }
    p^{J_1+\frac{r}{2}}q^{J_2+\frac{r}{2}}v_{\alpha}^{Q_{\alpha}}\,.
\end{equation}
In \eqref{SCIdef}, $J_{1,2}$ are the angular momenta of $S^3$, $r$ is the superconformal R-charge and $Q_\alpha$ are the generators of an eventual global symmetry of the theory.
The combinations $J_{1,2}+\frac{r}{2}$ are chosen to commute with $\mathcal{Q}$ and thus the SCI counts only those states that are annihilated by this supercharge.
Lastly, $p$, $q$ and $v_\alpha$ are fugacities associated to each symmetry.

The SCI can be rewritten as a matrix integral.
This can be obtained by exploiting its invariance under continuous deformations and evaluating it in the weak (gauge) coupling regime \cite{Kinney:2005ej,Sundborg:1999ue,Aharony:2003sx}.
Alternatively one recognizes the SCI as a supersymmetric partition function of the theory on $S^1 \times S^3$ and applies the localization techinique to compute it (modulo an overall phase due to the supersymmetric Casimir energy) \cite{Pestun:2007rz,Closset:2013vra,Assel:2014paa}.
In any case we end up with an integral representation of the SCI.
In what follows we restrict to the case of equal angular momenta, $p=q$, because this assumption holds for all the cases that we studied in this work.

Consider a 4d $\mathcal{N}=1$ gauge theory with semisimple gauge group $G$, global symmetry group $G_F$ and $\UU(1)_R$ R-symmetry.
The matter content consists of $n_\chi$ chiral multiplets in the representations $\mathfrak{R}_a$ of $G$, with weights $\rho_a$,  $\mathfrak{R}_F$ of $G_F$, with weights $\omega_a$, and with R-charges $r_a$. 
Then the integral formula for the SCI reads
\begin{equation}
    \label{SCInt}
    \mathcal{I}(v,q) = 
    \kappa \oint_{\numberset{T}^{\mathrm{rk}G}}
    \prod_{i=1}^{\mathrm{rk}G} \frac{\dd{z_i}}{2 \pi \mi z_i}\,
    \mathcal{Z}(z;v,q)\,,
\end{equation}
where
\begin{equation}
    \label{integrand}
    \kappa = 
    \frac{ (q;q)_{\infty}^{2\mathrm{rk}G} }{|\mathcal{W}|}\,,
    \qquad
    \mathcal{Z}(z;v,q) = 
    \frac
    {\prod_{a=1}^{n_\chi} \prod_{\rho_a \in \mathfrak{R}_a} \Gamma\left( q^{r_a} z^{\rho_a} v^{\omega_a}; q \right)}
    {\prod_{\alpha \in \Delta} \Gamma\left(z^\alpha;q\right)}\,.
\end{equation}
The integral contour is the maximal torus $\numberset{T}^{\mathrm{rk}G}$ of $G$ and $|\mathcal{W}|$ is the cardinality of the Weyl group of $G$.
The $q$-Pochhammer symbol $(p;q)_\infty$ and the elliptic gamma function $\Gamma(z;p,q)$ are 
defined\footnote{Notice that, since the angular momenta are equal, we dropped, for shortness, the last argument of the gamma functions. We kept this notation in the rest of the paper. We also introduced the notation
\begin{equation}
    z^{\rho_a}\equiv\prod_{i=1}^{\mathrm{rk}G} z_i^{\rho_a^i}\,,
    \quad
    v^{\omega_a}\equiv\prod_{i=1}^{\mathrm{rk}G_F} v_\alpha^{\omega_a^\alpha}\,.
\end{equation}
}
in \eqref{def:pochh} and \eqref{def:gammafug}.
From \eqref{integrand} we can distinguish the contribution of the matter fields in the numerator and the contribution of the gauge vector multiplet in the denominator where the product is taken over the set of roots $\Delta$ of $G$.
The representation \eqref{SCInt} is valid in the region
\begin{equation}
    \label{SCIdom}
    |q|<|q^{r_a}v^{\omega_a}|<1\,,
\end{equation}
where the elliptic functions in the integrand admit a plethystic expansion.
Eventually one perform an analytical continuation outside \eqref{SCIdom} after computing the integral.

The Bethe Ansatz approach rewrites the integral \eqref{SCInt} as a sum.
This is achieved by introducing a set of operators, dubbed Bethe Ansatz operators (BAOs): for any $i=1,\dots,\mathrm{rk}G$, we have
\begin{equation}
    \label{baos}
    Q_i(u;\Delta,\omega) \equiv 
    \prod_{a=1}^{n_\chi} \prod_{\rho_a \in \mathfrak{R}_a}
    P\left(\rho_a(u)+\Delta_a; \omega\right)^{\rho_a^i},
    \quad
    P(u;\omega) \equiv
    \frac{\ee^{ \pi \mi \left(u-\frac{u^2}{\omega}\right)}}
    {\theta_0(u;\omega)}\,,
\end{equation}
where the theta function $\theta_0$ is defined in \eqref{def:theta0} and we moved from fugacities to chemical potentials
\begin{equation}
    z_i=\ee^{2 \pi \mi u_i},
    \qquad
    q=\ee^{2 \pi \mi \omega},
    \qquad
    v_\alpha=\ee^{2 \pi \mi \xi_\alpha}.
\end{equation}
The chemical potential $\Delta_a$ is defined as
\begin{equation}
    \Delta_a \equiv \omega_a(\xi) + r_a \omega\,,
    \qquad
    y_a = \ee^{2 \pi \mi \Delta_a},
\end{equation}
with the constraint
\begin{equation}
    \sum_{a \in A} \Delta_a = 2 \omega\,,
\end{equation}
which reflects the invariance under flavor and R-symmetry of each term, here labelled by $A$, of the superpotential of the theory.
As shown in \cite{Benini:2018mlo}, the BAOs \eqref{baos} have the property to shift the argument of the integrand \eqref{integrand}
\begin{equation}
    Q_i(u;\Delta,\omega)\mathcal{Z}(u;\Delta,\omega) = 
    \mathcal{Z}(u-\delta_i \omega;\Delta,\omega)\,,
\end{equation}
therefore they can be used to recast the integral \eqref{SCInt} as
\begin{equation}
    \label{baf0}
    \mathcal{I}(\Delta,\omega) = 
    \kappa \int_{\partial\mathcal{A}} \prod_{i=1}^{\mathrm{rk}G}\mathrm{d}u_i\, \frac{\mathcal{Z}(u;\Delta,\omega)}{\prod_{i=1}^{\mathrm{rk}G}\left[1-Q_i(u;\Delta,\omega)\right]}\,,
\end{equation}
where the region $\partial \mathcal{A}$ is the boundary of the annulus
\begin{equation}
    \label{annulus}
    \mathcal{A} = \left\{
        u \in \C^{\mathrm{rk}G} \,|\,
        \Re u_i \in [0,1]\,,
        -\Im \omega < \Im u_i < 0\,,
        \forall \,i=1,\dots,\mathrm{rk}G
    \right\}\,.
\end{equation}
The integrand $\mathcal{Z}$ now is expressed in terms of chemical potentials, hence, for greater clarity, we emphasize that it is written as
\begin{equation}
    \label{integrand2}
    \mathcal{Z}(u;\Delta,\omega) =
    \frac{
        \prod_{a=1}^{n_\chi}
        \prod_{\rho_a \in \mathfrak{R}_a}
        \widetilde{\Gamma}(\rho_a(u)+\Delta_a;\omega)
    }{
        \prod_{\alpha \in \Delta}
        \widetilde{\Gamma}(\alpha(u);\omega)
    }\,,
\end{equation}
where the function $\widetilde \Gamma(u;\omega)$ is defined in \eqref{def:gamma}.
The integral \eqref{baf0} is computed by the residue theorem and it was proven in \cite{Benini:2018mlo} that the contributing poles come from the zeros of the denominator, hence the solutions of 
\begin{equation}
    \label{baes}
    Q_i(u;\Delta,\omega)=1\,,
    \qquad
    \forall \, i=1,\dots,\mathrm{rk}G\,,
\end{equation}
which are referred to as Bethe Ansatz equations (BAEs).
The invariance of the BAOs \eqref{baos} under the shifts $u_i \mapsto u_i + m + n \omega$ implies that actually there are infinite solutions of the BAEs \eqref{baes}.
However, one can collect them into a finite number of equivalence classes defined on a complex torus with modular parameter $\omega$, one for each $i$:
\begin{equation}
    \label{eqclass}
    u_i \sim u_i + 1 \sim u_i + \omega\,, 
    \qquad
    \forall \,i=1,\dots,\mathrm{rk}G
\end{equation}
Moreover, among all the solutions of \eqref{baes}, there are some of them that do not contribute to the integral \eqref{baf0}.
As discussed in the \textbf{Appendix C} of \cite{Benini:2018mlo}, these non contributing solutions are those ones that are fixed by a non trivial element of the Weyl group $\mathcal{W}$ of $G$.
These arguments lead to the following set of the contributing solutions
\begin{equation}
    \mathfrak{M}_{\text{BAE}} \equiv
    \left\{\,
        [\hat u] \in \mathcal{A} \,|\,
        Q_i(\hat{u};\Delta,\omega) = 1\,,
        w \cdot [\hat{u}] \ne [\hat{u}]\,,
        \forall \, i=1,\dots,\mathrm{rk}G\,,
        \forall \, 1\ne w \in \mathcal{W}\,
    \right\},
\end{equation} 
and thus the integral \eqref{baf0} becomes
\begin{equation}
    \label{baf}
    \mathcal{I}(\Delta,\omega) =
    \kappa \sum_{\hat{u} \in \mathfrak{M}_{\mathrm{BAE}}} 
    \mathcal{Z}(\hat{u};\Delta,\omega) \, \mathcal{H}(\hat{u};\Delta,\omega)^{-1}\,,
\end{equation}
where the Jacobian
\begin{equation}
    \mathcal{H}(u;\Delta,\omega) = 
    \det_{ij} \left(
        \frac{1}{2 \pi \mi} 
        \frac{\partial Q_i}{\partial u_j}
    \right),
\end{equation}
comes from the change of variables in the integral \eqref{baf0}.
The equation \eqref{baf} is the BA formula.

Whenever one is able to solve the BAEs, the problem of computing the SCI is translated into the evaluation of the sum \eqref{baf} on the solutions in $\mathfrak{M}_{\text{BAE}}$.
However, in the derivation of this result there is the assumption that the BAEs have only \emph{discrete} solutions, \emph{i.e.} isolated zeros.
It turns out that actually this is not always the case.
As we mentioned in the introduction, even in the low rank analysis of the case $G=\SU(N)$ with $N>2$, one finds that these discrete zeros are not enough to reproduce the SCI via \eqref{baf} \cite{Benini:2020gjh}.
There is a new set of solutions that describes a codimension one surface in the space of holonomies, with $\mathcal{H}=0$, dubbed \emph{continuous} solutions.
Therefore, the BA formula \eqref{baf} requires a generalization to include them into the sum.
Such generalization was recently proposed in \cite{Cabo-Bizet:2024kfe} where the authors evaluated the contribution of the continuous solutions for $\SU(3)$ and verified that the combination of their result with the contribution of the discrete roots reproduces the SCI at least at order $\order{1}$.
However, in this work, we discuss two cases in which  the BA formula \eqref{baf} holds in its original form, just involving the discrete solutions.

\section{The conifold theory}
\label{conisec}

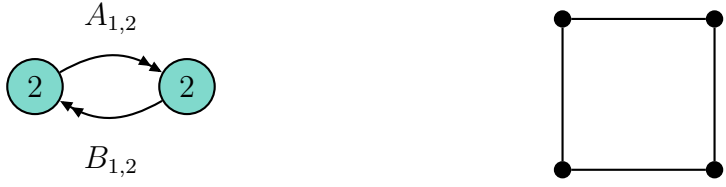
\begin{figure}
    \centering
    \begin{minipage}[b]{0.45\linewidth}
        \centering
        \begin{tikzpicture}[
            box/.style={rectangle, draw, thick},
        ]
        \pgfmathsetmacro{\x}{1}
        \pgfmathsetmacro{\y}{1}
        \node[fill=SUcol,circle,draw, thick] (l) at (-\x, 0) {$2$};
        \node[fill=SUcol,circle,draw, thick] (r) at (\x, 0) {$2$};
        \node at (0, \y-0.1) {$A_{1,2}$};
        \node at (0, -\y) {$B_{1,2}$};
        \draw[->>,  thick, >={Triangle[angle=45:5pt]}] (l) to[out=30, in=150, looseness=1.1] (r);
        \draw[->>,  thick, >={Triangle[angle=45:5pt]}] (r) to[out=-150, in=-30, looseness=1.1] (l);
        \end{tikzpicture}
    \end{minipage}
    \begin{minipage}[b]{0.45\linewidth}
        \centering
        \begin{tikzpicture}[
            box/.style={rectangle, draw, thick},
        ]
        \pgfmathsetmacro{\x}{1}
        \node[circle, fill=black, draw=black, line width=0.3mm, inner sep=2pt] (dot1) at (-\x, -\x) {};
        \node[circle, fill=black, draw=black, line width=0.3mm, inner sep=2pt] (dot2) at (\x, -\x) {};
        \node[circle, fill=black, draw=black, line width=0.3mm, inner sep=2pt] (dot3) at (\x, \x) {};
        \node[circle, fill=black, draw=black, line width=0.3mm, inner sep=2pt] (dot4) at (-\x, \x) {};
        \node[draw=none, fill=none] (invisibile) at (0,-1.1) {};
        \draw[-,  thick] (dot1) -- (dot2);
        \draw[-,  thick] (dot2) -- (dot3);
        \draw[-,  thick] (dot3) -- (dot4);
        \draw[-,  thick] (dot4) -- (dot1);
        \end{tikzpicture}
    \end{minipage}
    \caption{Quiver and toric diagrams of the conifold theory.}
    \label{conifig}
\end{figure}

The conifold theory is the theory living on a stack of $N$ D$3$-branes probing the tip of the conical singularity $xy-zt = 0$.
It is holographically dual to an $\mathrm{AdS}_5 \times T^{1,1}$ background, where the conifold is $T^{1,1} = \SU(2) \times \SU(2)/\UU(1)$. 
The theory has supersymmetry $\mathcal{N}=1$, the gauge group is $\SU(N)\times\SU(N)$ and there are two pairs of bifundamental chiral multiplets, precisely $A_1,A_2$ and $B_1,B_2$ that transform in the representations $(N,\bar{N})$ and $(\bar{N},N)$ respectively. 
The quiver and the toric diagram are shown in Figure \ref{conifig} and the superpotential is 
\begin{equation}
    W \sim \epsilon^{ij}\epsilon^{kl}
    \Tr A_i B_k A_j B_l\,.
\end{equation}
The global symmetry group is made up of a $\UU(1)_R$ R-symmetry, a $\UU(1)_B$ baryonic symmetry and two $\SU(2)$ flavor symmetries.
We summarize a charge assignation in the table below.

\[
\begin{array}{ccccc}
\toprule
 & \UU(1)_R & \UU(1)_B & \UU(1)_1 & \UU(1)_2 \\
\midrule
A_1 & \frac{1}{2} & 1 & 1 & 1 \\
A_2 & \frac{1}{2} & 1 & -1 & -1 \\
B_1 & \frac{1}{2} & -1 & 1 & -1 \\
B_2 & \frac{1}{2} & -1 & -1 & 1 \\
\bottomrule
\end{array}
\]
\\
Notice that, instead of the two flavor $\SU(2)$, we used combinations of their Cartans and denoted them by $\UU(1)_1, \UU(1)_2$.
However, it will be more useful to consider another parametrization.
Given the usual definitions for fugacities and chemical potentials associated to these symmetries
\begin{equation}
    p=q=\ee^{2\pi \mi \omega}\,, \quad 
    v_{1}=\ee^{2 \pi \mi \xi_{1}}\,, \quad
    v_{2}=\ee^{2 \pi \mi \xi_{2}}\,, \quad
    v_B = \ee^{2 \pi \mi \xi_B}\,,
\end{equation}
we define
\begin{align}
    &\Delta_1 = \xi_1 - \xi_2 - \xi_B + \tfrac{\omega}{2}\,,
    \quad
    &&\Delta_2 = -\xi_1 + \xi_2 - \xi_B + \tfrac{\omega}{2}\,,\\
    &\Delta_3 = \xi_1 + \xi_2 + \xi_B + \tfrac{\omega}{2}\,,
    \quad
    &&\Delta_4 = -\xi_1 - \xi_2 + \xi_B + \tfrac{\omega}{2}\,
\end{align} 
and the relative fugacities
\begin{equation}
    \label{yfug}
    y_a = \ee^{2 \pi \mi \Delta_a}\,, \quad a=1,2,3,4\,.
\end{equation}
These new chemical potentials are not independent: they satisfy the constraint
\begin{equation}
    \label{costr}
    \Delta_1+\Delta_2+\Delta_3+\Delta_4=2\omega\,,
\end{equation}
which reflects the invariance of the theory under flavor and R-symmetry.
Finally, for $N=2$ and equal angular momenta, the integral formula \eqref{SCInt} of the SCI for this theory reads
\begin{equation}
    \mathcal{I} = \frac{(q;q)_{\infty}^4}{4} \oint 
    \frac{\dd{z_1}}{2 \pi \mi z_1} \frac{\dd{z_2}}{2 \pi \mi z_2}
    \frac{
        \prod_{a=1}^{4}
        \Gamma\left( 
            z_1^{\pm1}z_2^{\pm1} y_a ; q 
        \right)
        \Gamma\left( 
            z_1^{\pm1}z_2^{\mp1} y_a ; q 
        \right)
    }{
        \Gamma\left( 
            z_1^{\pm2} ; q 
        \right)
        \Gamma\left( 
            z_2^{\pm2} ; q 
        \right)
    }\,.
\end{equation}

Now we apply the BA approach to such theory. 
We first show how it is possible to explicitly solve a sector of the BAEs, and then we use this result to compute the SCI.
Following the general formula \eqref{baos}, we have that, for each of the four bifundamental fields, the weights are
\begin{equation}
    \rho \in \left\{
    \left(1,1\right),\left(1,-1\right),\left(-1,1\right),\left(-1,-1\right)
    \right\},
\end{equation}
and we have one BAO for each node of the quiver:
\begin{equation}
    \begin{aligned}
        &Q^{(1)} = \prod_{\Delta \in \delta}
        \frac{P(u_1+u_2+\Delta)}{P(-u_1+u_2+\Delta)}
        \frac{P(u_1-u_2+\Delta)}{P(-u_1-u_2+\Delta)}=1\,,\\
        &Q^{(2)} = \prod_{\Delta \in \delta}
        \frac{P(u_1+u_2+\Delta)}{P(u_1-u_2+\Delta)}
        \frac{P(-u_1+u_2+\Delta)}{P(-u_1-u_2+\Delta)}=1\,,
    \end{aligned}
\end{equation}
where the set $\delta=\{\Delta_1,\Delta_2,\Delta_3,\Delta_4\}$ has been introduced for shortness.
At this point we change variables, moving to sum and difference of gauge holonomies
\begin{equation}
    \label{varcha}
    x = u_1 + u_2 \,, \quad y = u_1 - u_2\,,
\end{equation}
and we introduce two new BAOs as the product and the ratio of the original ones 
\begin{equation}
    Q_1 = \prod_{\Delta \in \delta} 
    \left[
        \frac{P(x+\Delta)}{P(-x+\Delta)}
    \right]^2=1\,,
    \quad
    Q_2 = \prod_{\Delta \in \delta} 
    \left[
        \frac{P(y+\Delta)}{P(-y+\Delta)}
    \right]^2=1\,.
\end{equation}
We use the definition \eqref{baos}, the relation among theta functions \eqref{def:theta} and the constraint \eqref{costr} to rewrite the BAEs as
\begin{equation}
    \label{conibae0}
    Q_1 = \prod_{\Delta \in \bar \delta} 
    \left[\frac{\theta_1(-x+\Delta)}{\theta_1(x+\Delta)}\right]^2=1\,,
    \quad
    Q_2 = \prod_{\Delta \in \bar \delta} 
    \left[\frac{\theta_1(-y+\Delta)}{\theta_1(y+\Delta)}\right]^2=1\,,
\end{equation}
where now the products run over the set $\bar \delta = \{\Delta_1,\Delta_2,\Delta_3,-\Delta_1-\Delta_2-\Delta_3\}$.
The two equations are equal, thus we focus just on the first one.
It is more convenient to consider the square root of the equation, so we have  
\begin{equation}
    \label{conibae}
    \prod_{\Delta \in \bar \delta} 
    \frac{\theta_1(x+\Delta)}{\theta_1(-x+\Delta)}=\pm1\,.
\end{equation}
Firstly, we notice that $x=0$ is a solution of the first sector (+).
Now, supposing that $x \ne 0$, we exploit the techiniques of \cite{Benini:2021ano} to procede in the proof.
Recalling the parity of $\theta_1$, \eqref{theta:even}, we multiply \eqref{conibae} by
\begin{equation}
    1=
    \frac
    {\theta_1(x)\theta_1(x-\Delta_1-\Delta_2)}
    {\theta_1(-x)\theta_1(-x-\Delta_1-\Delta_2)}
    \frac
    {\theta_1(x)\theta_1(x+\Delta_1+\Delta_2)}
    {\theta_1(-x)\theta_1(-x+\Delta_1+\Delta_2)}\,,
\end{equation}
so we rewrite the BAE as
\begin{equation}
    \label{conibae2}
    \frac
    {f(x;\Delta)g(x;\Delta) \mp f(-x;\Delta)g(-x;\Delta)}{f(-x;\Delta)g(-x;\Delta)}
    =0\,,
\end{equation}
where we introduced the auxiliary functions
\begin{equation}
    \begin{aligned}
        &f(x;\Delta) \equiv
        \theta_1(x)\theta_1(x+\Delta_1)\theta_1(x+\Delta_2)\theta_1(x-\Delta_1-\Delta_2)\,,\\
        &g(x;\Delta) \equiv
        \theta_1(x)\theta_1(x+\Delta_3)\theta_1(x+\Delta_1+\Delta_2)\theta_1(x-\Delta_1-\Delta_2-\Delta_3)\,.
    \end{aligned}
\end{equation}
These two functions can be rearranged using the Riemann identity \eqref{theta:riemann} 
\begin{align}
    &f(x;\Delta) = \sum_{r=1}^{4}c_r \theta_r(2x)\,,
    \quad
    &&c_r = \frac{1}{2}(-1)^{1+\delta_{r3}} 
    \theta_1(\Delta_1)\theta_1(\Delta_2)\theta_1(\Delta_1+\Delta_2)\,,\\
    &g(x;\Delta) = \sum_{r=1}^{4}d_r \theta_r(2x)\,,
    \quad
    &&d_r = \frac{1}{2}(-1)^{1+\delta_{r3}} 
    \theta_1(\Delta_3)\theta_1(\Delta_1+\Delta_2)\theta_1(\Delta_1+\Delta_2+\Delta_3)\,,\nonumber
\end{align}
and then, beacuse of the parity of theta functions \eqref{theta:even}, the product $fg$ becomes
\begin{equation}
    fg = 
    \left[\theta_1(2x) \sum_{r=2}^4\left(c_1 d_r + c_r d_1\right) \theta_r(2x)\right] + 
    \left[ c_1 d_1 \theta_1(2x)^2 +
    \sum_{r,s=2}^{4} c_r d_r \theta_r(2x)\theta_s(2x)\right],
\end{equation}
where we isolated the odd and the even parts.
If we restrict to the first sector of \eqref{conibae2}, we finally obtain 
\begin{equation}
    \label{conibae3}
    \frac{\theta_1(2x) h(x;\Delta)}{f(-x;\Delta)g(-x;\Delta)}=0\,,
\end{equation}
where we introduced another auxiliary function
\begin{equation}
    h(x;\Delta) \equiv \sum_{r=2}^4\left(c_1 d_r + c_r d_1\right) \theta_r(2x)\,.
\end{equation}
We thus found something very similar to the cases of $\mathcal{N}=4$ SYM with gauge group $\SU(2)$ and $\SU(3)$ \cite{Benini:2021ano}.
Here we have a continuous family of solutions given by
\begin{equation}
\label{conicont}
h(x;\Delta) = 0\,,
\end{equation} 
and a discrete family of solutions 
\begin{equation}
\label{conidiscr}
\theta_1(2x) = 0\,.
\end{equation} 
Once we have the zeros of both these equations, we have to remove all of them that are also zeros of the denominator. 
Therefore, from \eqref{theta:zeros} we obtain that the discrete solutions of \eqref{conibae3} are
\begin{equation}
\label{conisol}
 x = \frac{1}{2} ( m + n \omega )\,,  \quad \forall \, m,n \in \Z, \text{ with $m$,\,$n$ not both even}\,.
\end{equation}
The BAEs \eqref{conibae} are invariant under 
\begin{equation}
    x \mapsto x+1\,, \quad
    x \mapsto x+\omega\,, \quad
    x \mapsto -x\,,
\end{equation}
thus the distinct solutions of \eqref{conibae3} are the representatives of the equivalence class
\begin{equation}
    \label{claseq}
    x \sim x+1 \sim x+\omega \sim -x\,,
\end{equation}
and we conclude that 
\begin{equation}
    \label{conicomb}
    x \in \left\{ 0,\,\frac{1}{2}, \, \frac{\omega}{2}, \, \frac{1+\omega}{2}\right\},
\end{equation}
where we added the solution $x=0$, previously discussed.
All of this holds also for the other holonomy $y$.
Going back to the original holonomies $u_{1,2}$, \eqref{varcha}, we select, among all the possible combinations, the ones that contribute to the SCI, following the arguments discussed in Section \ref{barew}.
All those solutions that are fixed by non trivial element of the Weyl group do not contribute to the SCI via the BA formula \eqref{baf} \cite{Benini:2018mlo}. 
Here the Weyl group acts on the holonomies by sending
\begin{equation}
    u_1 \mapsto -u_1 \quad \text{or} \quad u_2 \mapsto -u_2\,,
\end{equation}
hence every solution that has a zero must be discarded.
Secondly, the contributing solutions can be chosen, for each direction, into a complex torus of modular parameter $\omega$ \cite{Benini:2018mlo}.
Lastly, since the BA formula only depends, in a symmetric way, on the combinations $x$ and $y$, we can group into a single equivalence class all those solutions that combine to give the same representative of the couple $(x,y)$ in the equivalence class \eqref{claseq}.
All these considerations lead us to conclude that the set of inequivalent discrete solutions of the first sector of \eqref{conibae} is
\begin{equation}
    \label{conimiracle}
    \mathfrak{M}_{\mathrm{BAE}} = \left\{
    \left(\tfrac{1}{4},\tfrac{1}{4}\right),
    \left(\tfrac{\omega}{4},\tfrac{\omega}{4}\right),
    \left(\tfrac{1+\omega}{4},\tfrac{1+\omega}{4}\right),\left(\tfrac{1+\omega}{4},\tfrac{1+3\omega}{4}\right),
    \left(\tfrac{2+\omega}{4},\tfrac{\omega}{4}\right),
    \left(\tfrac{1+2\omega}{4},\tfrac{1}{4}\right)
    \right\}\,.
\end{equation}
Notice that the first three elements of $\mathfrak{M}_{\text{BAE}}$ are HL solutions \cite{Hong:2018viz}, generalized to the case of two $\SU(2)$ gauge nodes.
As we discuss in Appendix \ref{HLsol}, they were already known to be solutions of the BAEs, even in the context of multiple gauge nodes \cite{Lanir:2019abx}.
On the other hand the last three elements are not of this type and they come into play to complete the set of allowed and distinct ways, on the torus, to obtain \eqref{conicomb} by \eqref{varcha}.

As we mentioned above, this result a priori does not conclude the analysis of the BA approach because we should still find the continuous family of solutions of \eqref{conicont} and solve the second sector of \eqref{conibae2}.
However, it turns out that the solutions \eqref{conimiracle} are enough to compute the SCI and thus there is no need to pursue this analysis further.
To present this result we first discuss the multiplicity of each solution.
As pointed in \cite{Benini:2018mlo}, if two solutions are related by a Weyl transformation, then they are equivalent in the BA formula.
This is true even for those solutions that correspond to the same $(x,y)$ modulo the equivalence relation \eqref{claseq}.
All of this is translated into a multiplicity of $8$ for each element of \eqref{conimiracle}.
To be more explicit we show how it works for the first element of the set \eqref{conimiracle}: starting from it we can generate these eight equivalent elements $(u_1,u_2)$ inside the torus
\begin{equation}
    \left(\tfrac{1}{4},\tfrac{1}{4}\right), 
    \left(\tfrac{1}{4},\tfrac{3}{4}\right), 
    \left(\tfrac{3}{4},\tfrac{1}{4}\right), 
    \left(\tfrac{3}{4},\tfrac{3}{4}\right),
    \left(\tfrac{1+2\omega}{4},\tfrac{1+2\omega}{4}\right),
    \left(\tfrac{1+2\omega}{4},\tfrac{1+3\omega}{4}\right),
    \left(\tfrac{1+3\omega}{4},\tfrac{1+2\omega}{4}\right),
    \left(\tfrac{1+3\omega}{4},\tfrac{1+3\omega}{4}\right),
\end{equation}
that correspond to $(x,y)=\left(\frac{1}{2},0\right)$.
To evaluate the SCI we expand it for small $q$.
To this aim we introduce a new parametrization 
\begin{equation}
    q = t^2\,, \quad
    y_1 = a b t\,, \quad
    y_2 = \frac{a t}{b}\,, \quad
    y_3 = \frac{c t}{b}\,, \quad
    y_4 = \frac{t}{bc}\,,
\end{equation}
that will simplify the evalutation.
In the case of the conifold theory the contribution of each solution \eqref{conimiracle} is not well defined in the case of equal fugacities (here $a=b=c=1$).
Therefore a suitable parametrization is useful when we look for a high order expansion. 
However, as expected, this regards only the intermediate steps because their sum, thus the full SCI, is well defined also in this limit.
Once the parametrization is chosen, the check proceeds as follows: on one hand we expand the integral formula \eqref{SCInt} of the index for small $t$ and then we evaluate the remaining integral, on the other hand we evaluate the SCI via \eqref{baf} and finally we compare the two results.
We present the final result in the case of equal fugacities, for brevity, while the individual contributions and the SCI for any fugacities are reported in Appendix \ref{SCIexp}.
The evaluation of the matrix integral gives
\begin{equation}
    \mathcal{I}_{T^{1,1}}(t) = 
    1 + 10\,t^2 + 50\,t^4 + 200\,t^6 + \order{t^8}\,,
\end{equation}
and this coincides with what we find from tha BA approach
\begin{equation}
    8\sum_{\mu=1}^6
    \mathcal{I}_\mu(t) = \mathcal{I}_{T^{1,1}}(t) + \order{t^8}\,,
\end{equation}
where the label $\mu$ represents the $\mu$-th solution in $\mathfrak{M}_{\mathrm{BAE}}$ \eqref{conimiracle}.

\section{The suspended pinch point theory}
\label{sppsec}

\begin{figure}
    \centering
    \begin{minipage}[b]{0.45\linewidth}
        \centering
        \makebox[\textwidth][c]{
        \begin{tikzpicture}[
            box/.style={rectangle, draw, thick},
            ]
        \pgfmathsetmacro{\x}{1.5}
        \pgfmathsetmacro{\y}{1}
        \pgfmathsetmacro{\z}{1}
        \pgfmathsetmacro{\a}{0.8}
        \pgfmathsetmacro{\b}{1.6}
        \node[fill=SUcol,circle,draw, thick] (t) at (0, \z) {$2$};
        \node[fill=SUcol,circle,draw, thick] (l) at (-\x, -\y) {$2$};
        \node[fill=SUcol,circle,draw, thick] (r) at (\x, -\y) {$2$};
        \draw[<->,  thick, >={Triangle[angle=45:5pt]}] (l) -- node[below]{$X_{23},X_{32}$} (r);
        \draw[<->,  thick, >={Triangle[angle=45:5pt]}] (l) -- node[left]{$X_{13},X_{31}$} (t);
        \draw[<->,  thick, >={Triangle[angle=45:5pt]}] (r) -- node[right]{$X_{12},X_{21}$} (t);
        \node at (\a, \b) {$\varphi$};
        \draw[->,  thick, >={Triangle[angle=45:5pt]}, looseness=6] (t) to[out=120, in=60] (t);
        \end{tikzpicture}
        }
    \end{minipage}
    \begin{minipage}[b]{0.45\linewidth}
        \centering
        \makebox[\textwidth][c]{
        \begin{tikzpicture}[
            box/.style={rectangle, draw, thick},
        ]
        \pgfmathsetmacro{\x}{2}
        \node[circle, fill=black, draw=black, line width=0.3mm, inner sep=2pt] (dot1) at (2*\x, 0) {};
        \node[circle, fill=black, draw=black, line width=0.3mm, inner sep=2pt] (dot2) at (\x, 0) {};
        \node[circle, fill=black, draw=black, line width=0.3mm, inner sep=2pt] (dot3) at (0, 0) {};
        \node[circle, fill=black, draw=black, line width=0.3mm, inner sep=2pt] (dot4) at (0, \x) {};
        \node[circle, fill=black, draw=black, line width=0.3mm, inner sep=2pt] (dot5) at (\x, \x) {};
        \node[draw=none, fill=none] (invisibile) at (0,-0.5) {};
        \draw[ thick] (dot1) -- (dot2);
        \draw[ thick] (dot2) -- (dot3);
        \draw[ thick] (dot3) -- (dot4);
        \draw[ thick] (dot4) -- (dot5);
        \draw[ thick] (dot5) -- (dot1);
        \end{tikzpicture}
        }
    \end{minipage}
    \\
    \vspace{15pt}
    \begin{tabular}{ccccccc}
    \toprule
    $\varphi$ & $X_{12}$ & $X_{21}$ & $X_{23}$ & $X_{32}$ &$ X_{31}$ & $X_{13}$ \\
    \midrule
    $\Delta_1+\Delta_2$ & $\Delta_4$ & $\Delta_3+\Delta_5$ & $\Delta_2$ & $\Delta_1$ & $\Delta_4+\Delta_5$ & $\Delta_3$ \\
    \bottomrule
    \end{tabular}
    \vspace{10pt}
    \caption{Quiver and toric diagrams of the SPP theory and charge parametrization.}
    \label{sppfig}
\end{figure}
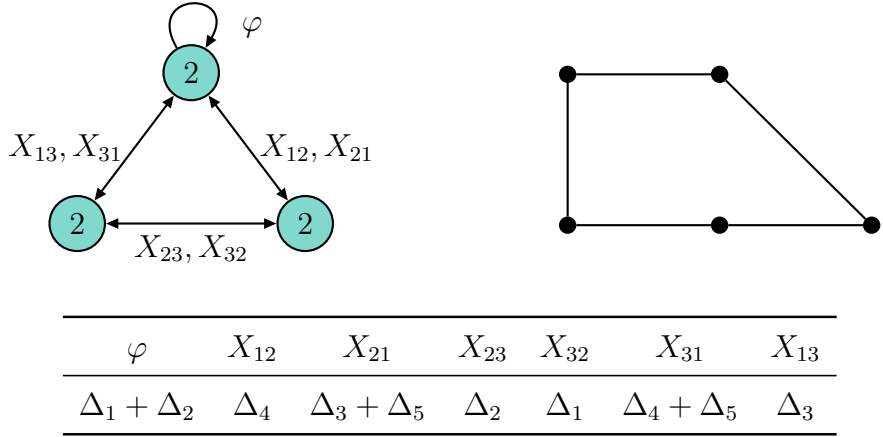

The suspended pinch point (SPP) theory corresponds to the near horizon limit of a stack of $N$ D3-branes probing the tip of the conical singularity $x^2y- wz = 0$. 
The supersymmetry is $\mathcal{N}=1$ and the gauge group is $\SU(N) \times \SU(N) \times \SU(N)$.
The matter content is constituted by an adjoint chiral $\varphi$ and six bifundamental chirals $X_{ij}$ with $i,j=1,2,3$ and $i \ne j$.
The quiver and the toric diagram are shown in Figure \ref{sppfig} and the superpotential is 
\begin{equation}
    W \sim \Tr \left\{ 
        \varphi X_{13}X_{31} - \varphi X_{12}X_{21} + 
        X_{21}X_{12}X_{23}X_{32} - X_{32}X_{23}X_{31}X_{13}
    \right\}\,.
\end{equation}
The global symmetries and a first charge assignation are summarized in the table below.

\[
\begin{array}{cccccc}
\toprule
 & \UU(1)_R & \UU(1)_1 & \UU(1)_2 & \UU(1)_3 & \UU(1)_4 \\
\midrule
\varphi & \frac{4}{5} & 1 & 1 & 0 & 0 \\
X_{12} & \frac{2}{5} & 0 & 0 & 0 & 1 \\
X_{21} & \frac{4}{5} & -1 & -1 & 0 & -1 \\
X_{13} & \frac{2}{5} & 0 & 0 & 1 & 0 \\
X_{31} & \frac{4}{5} & -1 & -1 & -1 & 0 \\
X_{23} & \frac{2}{5} & 0 & 1 & 0 & 0 \\
X_{32} & \frac{2}{5} & 1 & 0 & 0 & 0 \\
\bottomrule
\end{array}
\]
\vspace{15pt}

As before, given the chemical potentials $\xi_j$ associated to these $\UU(1)$ symmetries, we define the following combinations with the usual constraint
\begin{equation}
    \Delta_j = \xi_j + \frac{2}{5}\omega\,, 
    \qquad
    \Delta_5 = 2 \omega - \sum_{j=1}^{4} \Delta_j\,,
    \qquad
    \sum_{j=1}^{5}\Delta_j = 2 \omega\,. 
\end{equation}
The chemical potentials associated to each matter field, with respect to this parametrization, are summarized in Figure \ref{sppfig}.


We consider the case of $N=2$ and equal angular momenta, thus the integral formula of the SCI \eqref{SCInt} becomes
\begin{equation}
    \label{sppint}
    \mathcal{I} = \tilde \kappa \oint 
    \prod_{i=1}^3 \frac{\dd{z_i}}{2 \pi \mi z_i}
    \frac{
        \Gamma\left(z_1^{\pm2} \tilde y_\varphi;q\right)
        \prod_{a=1}^6 \prod_{i<j}^3
        \Gamma\left(z_i^{\pm1} z_j^{\pm1} \tilde y_a;q\right)
        \Gamma\left(z_i^{\pm1} z_j^{\mp1} \tilde y_a;q\right)
    }
    {\prod_{i=1}^3\Gamma\left(z_i^{\pm2};q\right)}\,,
\end{equation}
where 
\begin{equation}
    \tilde \kappa = \frac{(q;q)_{\infty}^6}{8}\,\Gamma(y_1y_2;q)\,,
\end{equation}
and
\begin{equation}
    \tilde y_\varphi = y_1 y_2\,, \quad
    \tilde y_1 = y_4\,, \quad
    \tilde y_2 = y_3 y_5\,, \quad
    \tilde y_3 = y_2\,, \quad
    \tilde y_4 = y_1\,, \quad
    \tilde y_5 = y_4 y_5\,, \quad
    \tilde y_6 = y_3\,, 
\end{equation}
with the fugacities $y_i$ defined as in the previous section in terms of $\Delta_i$ (see \eqref{yfug}).

Now we apply the BA approach to this theory.
Here the BAOs are 
\begin{equation}
\begin{aligned}
    Q^{(1)} &= 
    \prod_{i=2,3}
    \frac
    {P(u_1\pm u_i+\Delta_{1i})}
    {P(-u_1\pm u_i+\Delta_{1i})}
    \frac
    {P(u_1\pm u_i+\Delta_{i1})}
    {P(-u_1\pm u_i+\Delta_{i1})}
    \left[
        \frac
        {P(2u_1 + \Delta_{11})}
        {P(-2u_1 + \Delta_{11})}
    \right]^2 =1 \,, \\
    Q^{(2)} &= 
    \prod_{i=1,3}
    \frac
    {P(u_2\pm u_i+\Delta_{2i})}
    {P(-u_2\pm u_i+\Delta_{2i})}
    \frac
    {P(u_2\pm u_i+\Delta_{i2})}
    {P(-u_2\pm u_i+\Delta_{i2})} =1 \,,\\
    Q^{(3)} &= 
    \prod_{i=1,2}
    \frac
    {P(u_3\pm u_i+\Delta_{3i})}
    {P(-u_3\pm u_i+\Delta_{3i})}
    \frac
    {P(u_3\pm u_i+\Delta_{i3})}
    {P(-u_3\pm u_i+\Delta_{i3})} =1 \,,
\end{aligned}
\end{equation}
where, for shortness, we introduced the notation $\Delta_{ij}$ for the charge of each chiral $X_{ij}$:
\begin{align}
    &\Delta_{12} = \Delta_4\,,
    &&\Delta_{13} = \Delta_3\,,
    &&\Delta_{23} = \Delta_2\,, 
    &&\Delta_{11} = \Delta_1+\Delta_2\,,
    \nonumber\\
    &\Delta_{21} = \Delta_3+\Delta_5\,,
    &&\Delta_{31} = \Delta_4+\Delta_5\,,
    &&\Delta_{32} = \Delta_1\,.
    &&
\end{align}
As we did for the conifold, we express them in terms of $\theta_1$ using \eqref{baos} and \eqref{def:theta}, and we obtain
\begin{equation}
    \label{sppbaes}
    \begin{aligned}
        Q^{(1)} &= \ee^{32 \pi \mi u_1}
        \prod_{i=2,3}
        \frac
        {\theta_1(-u_1\pm u_i+\Delta_{1i})}
        {\theta_1(u_1\pm u_i+\Delta_{1i})}
        \frac
        {\theta_1(-u_1\pm u_i+\Delta_{i1})}
        {\theta_1(u_1\pm u_i+\Delta_{i1})}
        \left[
            \frac
            {\theta_1(-2u_1 + \Delta_{11})}
            {\theta_1(2u_1 + \Delta_{11})}
        \right]^2 =1 \,, \\
        Q^{(2)} &= \ee^{16 \pi \mi u_2}
        \prod_{i=1,3}
        \frac
        {\theta_1(-u_2\pm u_i+\Delta_{2i})}
        {\theta_1(u_2\pm u_i+\Delta_{2i})}
        \frac
        {\theta_1(-u_2\pm u_i+\Delta_{i2})}
        {\theta_1(u_2\pm u_i+\Delta_{i2})} =1 \,,\\
        Q^{(3)} &= \ee^{16 \pi \mi u_3}
        \prod_{i=1,2}
        \frac
        {\theta_1(-u_3\pm u_i+\Delta_{3i})}
        {\theta_1(u_3\pm u_i+\Delta_{3i})}        
        \frac
        {\theta_1(-u_3\pm u_i+\Delta_{i3})}
        {\theta_1(u_3\pm u_i+\Delta_{i3})} =1 \,.
    \end{aligned}
\end{equation}

We cannot yet solve \eqref{sppbaes}, differently from the case of the conifold above.
For this reason we looked for possible solutions with the following strategy.
First, as discussed in Appendix \ref{HLsol},  the HL solutions  solve the BAEs of SPP.
Then, we used the results of the conifold as guide to find other solutions.
In that case the combinations of holonomies appearing in the arguments of the $\theta_1$ functions in \eqref{conibae0} are sums and differences, and the solutions \eqref{conimiracle} are such that these combinations amount to 
\begin{equation}
\label{BEAHL}
    0\,, \frac{1}{2}\,, \frac{\omega}{2}\,, \frac{1+\omega}{2}\,.
\end{equation}
For the SPP model the combinations of holonomies in the BAEs \eqref{sppbaes} are sums, differences and multiplications by $2$.
Such operations applied on the HL solutions, which at this point are the only known solutions, give again \eqref{BEAHL} on the torus.
Therefore we guessed other solutions selecting all the possible inequivalent ways, on the torus, to obtain the terms in (\ref{BEAHL})  by acting on the holonomies with sums, differences and multiplications by $2$.
This led us to nine possibilities
\begin{samepage}
\begin{align}
    \label{sppmiracle}
    \mathfrak{M}_{\mathrm{BAE}} =
    \bigl\{
    &
    \left(\tfrac{1}{4},\tfrac{1}{4},\tfrac{1}{4}\right),
    \left(\tfrac{\omega}{4},\tfrac{\omega}{4},\tfrac{\omega}{4}\right),
    \left(\tfrac{1+\omega}{4},\tfrac{1+\omega}{4},\tfrac{1+\omega}{4}\right),
    \left(\tfrac{1+3\omega}{4},\tfrac{1+\omega}{4},\tfrac{1+\omega}{4}\right), \nonumber\\[5 pt]
    &
    \left(\tfrac{1+2\omega}{4},\tfrac{1}{4},\tfrac{1}{4}\right),
    \left(\tfrac{2+\omega}{4},\tfrac{\omega}{4},\tfrac{\omega}{4}\right),
    \left(\tfrac{1+\omega}{4},\tfrac{1+3\omega}{4},\tfrac{1+\omega}{4}
    \right),
    \left(\tfrac{1}{4},\tfrac{1+2\omega}{4},\tfrac{1}{4}\right),
    \left(\tfrac{\omega}{4},\tfrac{2+\omega}{4},\tfrac{\omega}{4}\right)
    \bigr\}\,.
\end{align}
\end{samepage}
Once we built a candidate for $\mathfrak{M}_{\text{BAE}}$, we checked that all its elements were solutions of \eqref{sppbaes} and then we evaluated their contribution to the SCI via \eqref{baf}.
This prescription does not guarantee that all the discrete solutions to  \eqref{sppbaes} have been obtained.
One could assume a more general ansatz (other rational coefficients, or even irrational) and check analytically, or at least numerically, that the BAEs are solved.
Here we did not broaden such analysis because we just aimed to reproduce the SCI, at least in a specific region of fugacities, and for this purpose we will show that the set \eqref{sppmiracle} is enough.

Using the same arguments discussed in Section \ref{conisec}, we obtain an overall multiplicity of $16$ for each element of \eqref{sppmiracle}: we have $8$ possibilities combining the first and the second holonomies and other $8$ combining the first and the third ones.
Moreover, for the last three elements we have an extra factor of $2$, which comes from the symmetry of the quiver, see Figure \ref{sppfig}: for each of them there is also another contributing solution with $u_2$ and $u_3$ exchanged.

As in the previous section, we compared the small $q=t^5$ expansion of the matrix integral \eqref{sppint} with the BA formula \eqref{baf} evaluated on \eqref{sppmiracle} with the relative multiplicities.
We chose the following parametrization for the fugacities:
\begin{equation}
    \label{paraspp}
    y_1= a t^{2}\,, \quad
    y_2= b t^{2}\,, \quad
    y_3= c t^{2}\,, \quad
    y_4= d t^{2}\,, \quad
    y_5= (abcd)^{-1} t^{2}\,.
\end{equation}
More precisely we first checked the matching at order $\order{1}$, thus we verified the cancellation of the terms with negative powers resulting from the expansion of the contributions of the each solution in \eqref{sppmiracle}.
However, it turned out that the two quantities coincides only in a specific region, that is
\begin{equation}
    \label{sppres0}
    16 \left.\left[\,
    \sum_{\mu=1}^{6}\mathcal{I}_{\mu}(t) + 
    2\sum_{\mu=7}^{9}\mathcal{I}_{\mu}(t)\,
    \right]\right|_{\mathcal{R}} = 1 + \order{t^{1/2}}\,,
\end{equation}
where $\mathcal{R}$ is the region defined by
\begin{equation}
    \label{sppslice}
    c=d \quad \text{and} \quad 
    \left(\,
        a=b^{-2}d^{-2}
        \quad \text{or} \quad
        b=a^{-2}d^{-2}\,
    \right)\,.
\end{equation}
In other words, outside the region \eqref{sppslice}, the discrete roots \eqref{sppmiracle} are not enough to reproduce the SCI.
One possible explanation of this circumstance could be that there are other solutions, discrete or continuous, whose contribution becomes non zero, and thus relevant, only when we go outside the region \eqref{sppslice}.
We refer the reader to the Appendix \ref{SCIexp} for the explicit results with different fugacities.

For SPP even the single contributions are well defined for equal fugacities, differently from the conifold theory, and this case is also inside the region $\mathcal{R}$ \eqref{sppslice}.
Therefore, in order to check \eqref{sppres0} beyond the leading order, we further restricted on this region of fugacities.
The small $q=t^5$ expansion of the matrix integral \eqref{sppint}, with $y_j=t^2$, gives
\begin{equation}
    \mathcal{I}_{\mathrm{SPP}}(t) = 1 + 5\,t^4 + 4\,t^6 + 20\,t^8 + 6\,t^9 + 18\,t^{10} + \order{t^{11}}\,.
\end{equation}
On the other hand we evaluated the BA formula and we obtained
\begin{equation}
    \label{sppres}
    16 \left[\,
    \sum_{\mu=1}^{6}\mathcal{I}_{\mu}(t) + 
    2\sum_{\mu=7}^{9}\mathcal{I}_{\mu}(t)\,
    \right] = \mathcal{I}_{\mathrm{SPP}}(t) + \order{t^{11}}\,,
\end{equation}
where the label $\mu$ represents the $\mu$-th solution in $\mathfrak{M}_{\mathrm{BAE}}$ \eqref{sppmiracle}.
Here we have shown the final result, while the single contributions to the SCI are deferred to the Appendix \ref{SCIexp}.


\section{Conclusions}

In this paper we studied the SCI for toric quiver $\SU(2)$ gauge theories with two and three gauge groups by using the BA approach.
A frequent problem occurring in this evaluation is related to the appearance of continuous solutions, that require a separated analysis in the evaluation of the index.
However, we found two models where the direct evaluation and the BA evaluation coincide by considering only discrete solutions.
In the first model, corresponding to the conifold, we have analytically solved one sector of the BAEs and evaluated the SCI for generic values of fugacities, showing a precise matching with the expected results. 
In the second model, corresponding to SPP,
we derived a broader set of solutions including the HL solutions and further elements, following the pattern observed in the conifold case.
Then we have evaluated the index on these solutions, finding that the BA formula does not reproduce the expected results for generic values of the flavor fugacities, but it matches with the direct evaluation by opportunely turning off some global symmetries.

Some comments are in order. First, observe that,  at least up to our knowledge, this behavior looks new. Indeed in other cases where discrete solutions do not reproduce the exact evaluations, like $\SU(N)$ $\mathcal{N}=4$ SYM, the divergent index evaluated on the discrete solutions to the BAEs does not have a well defined limit associated to turning off some flavor fugacities.
Here, in the case of the SPP model, we have found on the other hand that such limit is well defined and that it reproduces the exact evaluation. 
However, we have not found the exact matching with all the flavor fugacities turned on. 
It would be desirable to obtain such matching, either by finding new discrete solutions or  continuous solutions, evaluating the SCI in this second case along the recent discussion of \cite{Cabo-Bizet:2024kfe}.
Furthermore, beyond the models  shown in this note we have  scanned over all the toric quivers with two and three gauge groups, corresponding to $\mathbb{Z}_{2,3}$ orbifolds of $\mathbb{C}^3$. 
In such cases we have not found neither the behavior of $T^{1,1}$ nor of SPP, \emph{i.e.} we did not find any regime of charges that allows to reproduce the exact index with the BA approach, by evaluating the SCI using the solutions obtained with our prescription.
Similarly to the case of SPP, we cannot conclude with this approach neither if other discrete solutions contribute to the index in such cases nor if the missing contributions originate from continuous families. 
A last comment regards the analogy between the analysis of the BAEs for the conifold and for $\SU(3)$ $\mathcal{N}=4$ SYM done in \cite{Benini:2021ano}. 
The two sets of BAEs are solved in the two cases in a quite identical manner. 
Anyway, in our case, by evaluating the SCI on such solutions we reproduced the direct evaluation results, while in  \cite{Benini:2021ano} such a matching failed, and continuous solutions became necessary.
Motivated by the exact matching, here, we have not studied possible continuous solutions for the BAEs of the conifold, because even if such solutions exist they do not contribute to the SCI for $T^{1,1}$ with $\SU(2)\times \SU(2)$ gauge group.

\section*{Acknowledgments}
This work  has been supported in part by the Italian Ministero dell'Istruzione, Università e Ricerca (MIUR), in part by Istituto Nazionale di Fisica Nucleare (INFN) through the “Gauge Theories, Strings, Supergravity” (GSS) research project.

\appendix

\section{Special functions}
\label{app:specfunc}

We give the definitions of some special functions, then we list all their properties that we need in this work and fix the notation.
We use fugacities and chemical potentials related by
\begin{equation}
\label{def:fug}
z=\ee^{2 \pi \mi u}\,,
\qquad
p=\ee^{2 \pi \mi \sigma}\,,
\qquad
q=\ee^{2 \pi \mi \tau}\,,
\end{equation}
with $|p|, |q| < 1$.
The $q$-Pochhammer symbol is defined as 
\begin{equation}
\label{def:pochh}
(p;q)_{\infty} \equiv 
\prod_{n=0}^{\infty} \left( 1-p q^n \right)\,.
\end{equation}
The elliptic gamma function is defined as
\begin{equation}
\label{def:gammafug}
\Gamma(z;p,q) \equiv
\prod_{m,n=0}^{\infty} \frac{ 1 - p^{m+1}q^{n+1}z^{-1} }{ 1 - p^{m}q^{n}z }\,,
\end{equation}
but we sometimes refer to gamma as function of chemical potentials
\begin{equation}
\label{def:gamma}
\widetilde{\Gamma}(u;\tau,\sigma) \equiv
\Gamma( \ee^{2 \pi \mi u}; \ee^{2 \pi\mi \tau}, \ee^{2 \pi\mi \sigma} )\,.
\end{equation}
Then we have the theta function $\theta_0$ which is defined as
\begin{equation}
\label{def:theta0}
\theta_0(u;\omega) \equiv
(z;h)_{\infty} (h z^{-1};h)_{\infty}\,.
\end{equation}
Finally we have the following definitions for the Jacobi theta functions
\begin{align}
    \label{def:theta}
    &\theta_1(u;\omega) \equiv 
    \mi \ee^{ -\pi \mi \left( u - \frac{\omega}{4} \right) } (h;h)_{\infty} \theta_0(u;\omega)\,,
    \quad
    &&\theta_2(u;\omega) \equiv
    \ee^{ -\pi \mi \left( u - \frac{\omega}{4} \right) } (h;h)_{\infty} \theta_0\left(u + \tfrac{1}{2};\omega \right)\,, \nonumber\\
    &\theta_3(u;\omega) \equiv
    (h;h)_{\infty} \theta_0\left(u + \tfrac{1}{2} + \tfrac{\omega}{2};\omega \right)\,, 
    \quad
    &&\theta_4(u;\omega) \equiv
    (h;h)_{\infty} \theta_0\left(u + \tfrac{\omega}{2};\omega \right)\,.
\end{align}

We list some properties of these functions that we used in this work.
Firstly, there is the quasi-double-periodicity, $\forall \, m,n \in \Z$,
\begin{equation}
\label{theta:doubleper}
\begin{split}
\theta_0(u+m+n\omega;\omega) &= (-1)^{n} \ee^{-2 \pi \mi n u} e^{- \pi \mi n(n-1) \omega} \theta_0(u;\omega)\,,  \\
\theta_1(u+m+n\omega;\omega) &= (-1)^{m+n} \ee^{-2 \pi \mi n u} \ee^{- \pi \mi n^2 \omega} \theta_1(u;\omega)\,.
\end{split}
\end{equation}
and the evenness properties, $\forall \, r=1,2,3,4$,
\begin{equation}
\label{theta:even}
\theta_r(-u;\omega) = (-1)^{\delta_{r,1}} \theta_r(u;\omega)\,.
\end{equation}
From the quasi-double-periodicity \eqref{theta:doubleper} and the analogous identities for the other theta functions, shifting by periods or half-periods, we obtain their (first order) zeros \cite{Kharchev_2015}:
\begin{equation}
\label{theta:zeros}
\begin{split}
\theta_1(u;\omega)=0 \, : \qquad &u = m + n \omega\,,  \\
\theta_2(u;\omega)=0 \, : \qquad &u = m + \tfrac{1}{2} + n \omega\,,  \\
\theta_3(u;\omega)=0 \, : \qquad &u = m + \tfrac{1}{2} + \bigl( n+\tfrac{1}{2} \bigr)\omega\,,  \\
\theta_4(u;\omega)=0 \, : \qquad &u = m + \bigl( n+\tfrac{1}{2} \bigr)\omega\,, 
\end{split}
\end{equation}
where $m,n \in \Z$. 

Lastly, we use a five term Riemann identity \cite{Kharchev_2015}, which involves linear combinations of five products of four $\theta$ functions.
We simplify the notation by imposing that $\theta_r(u;\omega) \equiv \theta_r(u)$. 
Then we introduce another set of variables $W$, $X$, $Y$, $Z$, their dual counterparts
\begin{equation}
\begin{split}
&W' = \tfrac{1}{2}\bigl( -W+X+Y+Z \bigr)\,,  \\
&X' \,= \tfrac{1}{2}\bigl( W-X+Y+Z \bigr)\,,  \\
&Y' \;= \tfrac{1}{2}\bigl( W+X-Y+Z \bigr)\,,  \\
&Z' \;= \tfrac{1}{2}\bigl( W+X+Y-Z \bigr)
\end{split}
\end{equation}    
and the compact notation
\begin{equation}
\begin{split}
&[r] \;= \theta_r(W)\theta_r(X)\theta_r(Y)\theta_r(Z)\,,  \\
&[r]' = \theta_r(W')\theta_r(X')\theta_r(Y')\theta_r(Z')\,.
\end{split}
\end{equation}
In this language the identity that we need is 
\begin{equation}
\label{theta:riemann}
2\,[1]' = [1] + [2] - [3] + [4]\,.
\end{equation}

\section{The Hong-Liu solutions}
\label{HLsol}

In this section we explicitly verified that the HL solutions solve the BAEs \eqref{baes} in the cases of the conifold theory and the SPP theory with gauge group factors $\SU(N)$, for any $N$.
These solutions have been introduced in \cite{Hong:2018viz}, in the context of the BA approach applied to topologically twisted index of $\mathcal{N}=4$ SYM theory on $S^2 \times T^2$ with gauge group $\SU(N)$.
Subsequently, they were also considered in the case of SCI of $\mathcal{N}=4$ SYM theory on $S^3$ with gauge group $\SU(N)$ \cite{Benini:2018ywd,Benini:2021ano,Aharony:2021zkr}.
They were also mentioned in the context $\mathcal{N}=1$ toric theories in \cite{Lanir:2019abx}.

Here, to verify that they solve the BAEs \eqref{baes}, we will apply the argument given in \cite{Hong:2018viz}, generalizing what has been done in \cite{Lanir:2019abx} for the basic solution, which is a specific element of this family. 

Before proceding with the proof, it is more convenient to rewrite the BA approach in terms of constrained $\UU(N)$ gauge groups instead of $\SU(N)$.
We give a brief review of this change of formalism for a single gauge group $\SU(N)$. 
See \cite{Benini:2018ywd} for a comprehensive discussion about it.
However, this argument can be straightforwardly generalized to the case of multiple $\SU(N)$ factors..

The $\SU(N)$ holonomies are $u_1,\dots,u_{N-1}$, but we can describe them also in terms of the holonomies $u_1,\dots,u_{N}$ of $\UU(N)$, costrained by
\begin{equation}
    \label{sucostr}
    \sum_{i=1}^{N}u_i = 0  \mod \Z + \omega \Z\,.
\end{equation}
The modding is due to the fact that the holonomies are defined on the torus with modular parameter $\omega$.
Once we moved to the $\UU(N)$ language, we need to modify the BAEs because now we have one extra BAO, $Q_N$, but this one is not really independent on the others.
This can be achieved by introducing a sort of “Lagrange multiplier” $\lambda$ in the definition of the BAOs of $\UU(N)$, and this reflects the constraint among the $Q_i$ to reproduce the $\SU(N)$ behavior.
For example, for $\mathcal{N}=4$ SYM theory with gauge group $\SU(N)$, the BAOs \eqref{baos} become
\begin{equation}
    \label{ubaes}
    Q_i = \ee^{ 2 \pi \mi \lambda } \prod_{\Delta \in \delta} \prod_{j=1}^N 
    \frac{ P( u_{ij} + \Delta ; \omega ) }{ P( u_{ji} + \Delta ; \omega ) }\,,
    \qquad
    i=1,\dots,N\,,
\end{equation}
where $u_{ij} = u_i-u_j$, $\delta = \{\Delta_1,\Delta_2,\Delta_3\}$ and $P(u;\omega)$ is defined in \eqref{baos}.
Notice that also the weights here are translated in the language of $\UU(N)$, therefore for the adjoint representation they are
\begin{equation}
    \text{adj} \ni \rho \sim (\dots,1,\dots,-1,\dots)\,,
\end{equation}
where the dots are zeros.

In this formalism the HL solutions are 
\begin{equation}
    \label{hl}
    \{m,n,r\} : \quad
    u_{\hat{\jmath}\hat{\kappa}} = 
    \bar{u} + 
    \frac{\hat{\jmath}}{m} + 
    \frac{\hat{\kappa}}{n} \left( \omega + \frac{r}{m} \right) 
    \equiv 
    \bar{u} + \frac{ \hat{\jmath} + \hat{\kappa} \tilde{\omega}}{m}\,,
\end{equation}
where $m,n,r \in \Z \, : \, mn=N$, $r \in [0,n)$ and $\hat{\jmath}=0,\dots,m-1$, $\hat{\kappa}=0,\dots,n-1$.
The constant $\bar{u}$ is chosen to satisfy \eqref{sucostr}.
Now we describe the argument of \cite{Hong:2018viz} to show that \eqref{hl} solve \eqref{ubaes}.
First, rewrite \eqref{ubaes} in terms of the functions $\theta_1$ by \eqref{def:theta}.
Then using the constraint among the chemical potentials 
\begin{equation}
    \Delta_1+\Delta_2+\Delta_3=2\omega\,,
\end{equation}
and the quasi-double-periodicity \eqref{theta:doubleper} of $\theta_1$, the BAEs \eqref{ubaes} become 
\begin{equation}
    \label{baesc}
    Q_i = \ee^{ 2 \pi \mi \lambda } 
    \prod_{\Delta \in \bar{\delta}} \prod_{j=1}^N 
    \frac{\theta_1(u_{ji} + \Delta)}{\theta_1(u_{ij} + \Delta)}=1\,,
\end{equation}
where $\bar{\delta} = \{\Delta_1,\Delta_2,-\Delta_1-\Delta_2\}$ and $i=1,\dots,N$.
Now we evaluate \eqref{baesc} on \eqref{hl}. 
To this aim we split the product over $j$ as 
\begin{equation}
    \prod_{j=1}^N \longrightarrow \prod_{\hat{\jmath}=0}^{m-1}\prod_{\hat{\kappa}=0}^{n-1}\,,
\end{equation}
and the differences become
\begin{equation}
u_{ij} \longrightarrow u_{ \hat{\jmath} \hat{\jmath}_0 \hat{\kappa} \hat{\kappa}_0 } = \frac{1}{m} \left[ (\hat{\jmath} - \hat{\jmath}_0) + \tilde{\omega} (\hat{\kappa} - \hat{\kappa}_0) \right]\,.
\end{equation}
Then we use the identity (see also \cite{Hong:2018viz})
\begin{equation}
    \label{hlshift}
    \prod_{\hat{\jmath}=0}^{m-1}\prod_{\hat{\kappa}=0}^{n-1}
    \theta_1\left(
        \Delta \pm 
        \frac{1}{m} \left[
            (\hat{\jmath} - \hat{\jmath}_0) +
            \tilde{\omega} (\hat{\kappa} - \hat{\kappa}_0)
        \right]
    \right) = 
    \zeta
    \prod_{\hat{\jmath}=0}^{m-1}\prod_{\hat{\kappa}=0}^{n-1}
    \theta_1\left(
        \Delta \pm 
        \frac{\hat{\jmath} + \tilde{\omega} \hat{\kappa}}{m}
    \right)\,,
\end{equation}
where
\begin{equation}
    \label{coeffc}
    \zeta = 
    (-1)^{n \hat{\jmath}_0 + (m+r)\hat{\kappa}_0}
    \ee^{-\pi \mi m \hat{\kappa}_0 \omega}
    \ee^{\pm \pi \mi \hat{\kappa}_0
    \left[ 
        2 m \Delta \pm \tilde{\omega} (2n-1-\hat{\kappa}_0)
        \right]}\,,
\end{equation}
which can be obtained by the quasi-double-periodicity \eqref{theta:doubleper} of $\theta_1$.
Upon substituting \eqref{hlshift} into the BAOs in \eqref{baesc}, the coefficients $\zeta$ in each term cancel, yielding
\begin{equation}
    Q_{\hat{\jmath}_0 \hat{\kappa}_0} = 
    \ee^{2 \pi \mi \lambda} 
    \prod_{\hat{\jmath}=0}^{m-1}\prod_{\hat{\kappa}=0}^{n-1}
    \frac{
        \theta_1\left(
        \Delta - 
        \frac{\hat{\jmath} + \tilde{\omega} \hat{\kappa}}{m}
        \right)
    }{
        \theta_1\left(
        \Delta + 
        \frac{\hat{\jmath} + \tilde{\omega} \hat{\kappa}}{m}
        \right)
    }\,,
\end{equation}
where the right-hand side no longer depends on $\hat{\jmath}_0$ or $\hat{\kappa}_0$.
Finally, we are still free to choose $\lambda$, therefore, by requiring that
\begin{equation}
    \ee^{2 \pi \mi \lambda}
    \overset{!}{=}
    \prod_{\hat{\jmath}=0}^{m-1}\prod_{\hat{\kappa}=0}^{n-1}
    \frac{
        \theta_1\left(
        \Delta +
        \frac{\hat{\jmath} + \tilde{\omega} \hat{\kappa}}{m}
        \right)
    }{
        \theta_1\left(
        \Delta -
        \frac{\hat{\jmath} + \tilde{\omega} \hat{\kappa}}{m}
        \right)
    }\,,
\end{equation}
the BAEs are automatically solved.

In what follows we apply this argument to the conifold and SPP theories.
In the case of multiple $\SU(N)$ gauge factors the HL solutions are generalized as
\begin{equation}
    \label{hlt}
    \{m,n,r\} : \quad
    u_{\hat{\jmath}\hat{\kappa}}^{(\alpha)} = 
    \bar{u} + 
    \frac{\hat{\jmath}}{m} + 
    \frac{\hat{\kappa}}{n} \left( \omega + \frac{r}{m} \right) 
    \equiv 
    \bar{u} + \frac{ \hat{\jmath} + \hat{\kappa} \tilde{\omega}}{m}\,.
\end{equation}
Notice that these solutions are built by taking the same HL solution for each node.
Moreover we recall that in the $\UU(N)$ language the weights of the bifundamental representation are
\begin{equation}
\begin{aligned}
&(N, \bar{N}) \ni \rho \sim ( \underbrace{\dots,1,\dots}_N, \underbrace{\dots,-1,\dots}_N )\,,  \\
&(\bar{N},N) \ni \rho \sim ( \underbrace{\dots,-1,\dots}_N, \underbrace{\dots,1,\dots}_N )\,,
\end{aligned}
\end{equation}
where the dots are zeros. 

We start with the conifold theory with gauge group $\SU(N) \times \SU(N)$.
We used the same steps discussed above to rewrite the BAEs: the definition of $\theta_1$ \eqref{def:theta}, its quasi-double-periodicity \eqref{theta:doubleper} and the constraint
\begin{equation}
    \Delta_1+\Delta_2+\Delta_3+\Delta_4=2\omega.
\end{equation}
Combining all of this we get
\begin{equation}
    \label{eq:conibaes}
    \begin{aligned}
    Q_i^{(1)} &= 
    e^{ 2 \pi \mi \lambda^{(1)} }  \prod_{j=1}^N \, 
    \frac
    { \theta_1\left( u_{ji}^{(21)} +\Delta_3 \right) \, \theta_1\left( u_{ji}^{(21)} - \Delta_1 - \Delta_2 - \Delta_3 \right) }
    { \theta_1\left( u_{ij}^{(12)} +\Delta_1 \right) \, \theta_1\left( u_{ij}^{(12)} +\Delta_2 \right) } = 1\,,      \\
    Q_i^{(2)} &= 
    e^{ 2 \pi \mi \lambda^{(2)} } \prod_{j=1}^N \, 
    \frac
    { \theta_1\left( u_{ji}^{(12)} +\Delta_1 \right) \, \theta_1\left( u_{ji}^{(12)} +\Delta_2 \right) }
    { \theta_1\left( u_{ij}^{(21)} +\Delta_3 \right) \, \theta_1\left( u_{ij}^{(21)} - \Delta_1 - \Delta_2 - \Delta_3 \right) } = 1\,,
    \end{aligned}
\end{equation}
where $u_{ij}^{\alpha \beta} = u_i^{(\alpha)}-u_j^{(\beta)}$ and $i=1,\dots,N$.
We evaluate \eqref{eq:conibaes} on the HL solutions \eqref{hlt}, we use the relation \eqref{hlshift} and again all the terms depending on $\hat{\jmath}_0$ or $\hat{\kappa}_0$ cancel.
The suitable choice of the remaining Lagrange multipliers $\lambda^{(\alpha)}$ guarantees that the BAEs \eqref{eq:conibaes} are solved.

Now we consider the SPP theory with gauge group $\SU(N) \times \SU(N) \times \SU(N)$.
Also in this case we rewrite the BAEs using \eqref{def:theta}, \eqref{theta:doubleper} and the constraint
\begin{equation}
    \Delta_1+\Delta_2+\Delta_3+\Delta_4+\Delta_5=2\omega\,,
\end{equation}
and we obtain, for $i=1,\dots,N$,
\small{
\begin{equation}
\label{eq:sppbaes}
\begin{aligned}
    Q_i^{(1)} &= \ee^{2 \pi \mi \lambda^{(1)}} \prod_{j=1}^N 
    \frac{
        \theta_1\left( u_{ji}^{(11)} \! + \! \Delta_1 \! + \! \Delta_2 \right)
    }{
        \theta_1\left( u_{ij}^{(11)} \! + \! \Delta_1 \! + \! \Delta_2 \right)
    }
    \frac{
        \theta_1\left( u_{ji}^{(21)} \! - \! \Delta_1 \! - \! \Delta_2 \! - \! \Delta_4 \right)
    }{
        \theta_1\left( u_{ij}^{(12)} \! + \! \Delta_4 \right)
    }
    \frac{
        \theta_1\left( u_{ji}^{(31)} \! - \! \Delta_1 \! - \! \Delta_2 \! - \! \Delta_3 \right)
    }{
        \theta_1\left( u_{ij}^{(13)} \! + \! \Delta_3 \right)
    }\,, 
    \\
    Q_i^{(2)} &= \ee^{2 \pi \mi \lambda^{(2)}} \prod_{j=1}^N 
    \frac{
        \theta_1\left( u_{ji}^{(12)} \! + \! \Delta_4 \right)
    }{
        \theta_1\left( u_{ij}^{(21)} \! - \! \Delta_1 \! - \! \Delta_2 \! - \! \Delta_4 \right)
    }
    \frac{
        \theta_1\left( u_{ji}^{(32)} \! + \! \Delta_1 \right)
    }{
        \theta_1\left( u_{ij}^{(23)} \! + \! \Delta_2 \right)
    }\,,
    \\
    Q_i^{(3)} &= \ee^{2 \pi \mi \lambda^{(3)}} \prod_{j=1}^N 
    \frac{
        \theta_1\left( u_{ji}^{(13)} \! + \! \Delta_3 \right)
    }{
        \theta_1\left( u_{ij}^{(31)} \! - \! \Delta_1 \! - \! \Delta_2 \! - \! \Delta_3 \right)
    }
    \frac{
        \theta_1\left( u_{ji}^{(23)} \! + \! \Delta_2 \right)
    }{
        \theta_1\left( u_{ij}^{(32)} \! + \! \Delta_1 \right)
    }\,.
\end{aligned}
\end{equation}}
We evaluate \eqref{eq:sppbaes} on \eqref{hlt} and, using the relation \eqref{hlshift}, the same cancellations occur and the suitable choice of the Lagrange multipliers $\lambda^{(\alpha)}$ solves the BAEs \eqref{eq:sppbaes}.

\section{Series expansions of the SCI}
\label{SCIexp}

In this section we show explicity the contributions that we obtained in the expansion of the SCI via \eqref{baf}.

\subsection*{Conifold }
\label{coniapp}

We first list the “tachyonic” contributions of each discrete solution of the BAEs in \eqref{conimiracle}.
We focus on divergent terms to show explicitly the cancellations that occur among them.
Given the denominator
\begin{equation}
    D(a,b,c) = \frac{1}{4(a-c)^2 (a c-1)^2(b^2-1)^2} \equiv D\,,
\end{equation}
which is common to all the contributions, we have
\begin{align}
    &\mathcal{I}_1 = 
    - D \, a^2 b^2 c^2 \,\frac{1}{t^2}+\order{1}\,, \nonumber\\
    &
    \begin{aligned}
        \mathcal{I}_2 
        &=
        -\frac{D}{16}\,
        a^2 b^2 c^2\,\frac{1}{t^3} + \\
        &+ \frac{D}{16}\,
        \bigl[a b c (c (-(a^2+1) b^2+a^2+8 a b+1)+a (b^2-1) c^2+ a (b^2-1))\bigr]\,
        \frac{1}{t^2}+ \\
        &+\frac{D}{16}\,
        \bigl[a^2 (b^2-1) c^4+a^2 (b^2-1)+a c^3 ((a^2+1) b^4-4 (a^2+1) b^2+a^2 +\\
        &-6 a b^3+6 a b+1)+a c ((a^2+1) b^4-4 (a^2+1) b^2+a^2 -6 a b^3+6 a b+1) +\\
        &-c^2 (-6 a (a^2+1) b^3+6 a (a^2+1) b+4 a^2+(a^4+4 a^2+1) b^4 +\\
        &- (a^4-24 a^2+1) b^2)\bigr]\,
        \frac{1}{t}+\order{1}\,,
    \end{aligned}
    \nonumber\\
    &
    \begin{aligned}
        \mathcal{I}_3 
        &=
        \frac{D}{16}\,
        a^2 b^2 c^2\,\frac{1}{t^3}+\\
        &+\frac{D}{16}\,
        \bigl[abc (a^2 (b^2-1)c+a (-b^2 (c^2+1)+8 b c+c^2+1)+(b^2-1) c)\bigr]\,
        \frac{1}{t^2}+\\
        &+\frac{D}{16}\,
        \bigl[a^4 b^2(b^2-1)c^2-a^3 c (b^4 (c^2+1)-6 b^3 c-4 b^2 (c^2+1)+6 b c+c^2+ \\
        &+1)+a^2 (4 b^4 c^2-6 b^3 (c^3+c)-b^2 (c^4-24 c^2+1)+6 b (c^3+c)+c^4+ \\
        &+4 c^2+1)-a c (b^4 (c^2+1)-6 b^3 c-4 b^2 (c^2+1)+6 b c+c^2+1)+ \\
        &+b^2 (b^2-1) c^2\bigr]\,
        \frac{1}{t}+\order{1}\,,
    \end{aligned}
    \nonumber\\
    &
    \begin{aligned}
        \mathcal{I}_4
        &=
        \frac{D}{16}\,
        a^2 b^2 c^2\,\frac{1}{t^3}+ \\
        &+\frac{D}{16}\,
        \bigl[a b c (c (-(a^2+1) b^2+a^2+8 a b+1)+ a (b^2\!-\!1)c^2+a (b^2\!-\!1))\bigr]\,
        \frac{1}{t^2}+ \\
        &+\frac{D}{4}\,
        \bigl[a^4 b^2 (b^2-1)c^2 -a^3 c (b^4 (c^2+1)+6 b^3 c-4 b^2 (c^2+1)-6 b c+c^2+\\
        &+1)+a^2 (4 b^4 c^2+6 b^3 (c^3+c)-b^2 (c^4-24 c^2+1)-6 b (c^3+c)+c^4+\\
        &+4 c^2+1)-a c (b^4 (c^2+1)+6 b^3 c-4 b^2 (c^2+1)-6 b c+c^2+1)+\\
        &+b^2 (b^2-1) c^2\bigr]\,
        \frac{1}{t}+\order{1}\,,
    \end{aligned}
    \nonumber\\
    &
    \begin{aligned}
        \mathcal{I}_5 
        &= 
        -\frac{D}{16}\,
        a^2 b^2 c^2\,\frac{1}{t^3} + \\
        &+ \frac{D}{16}
        \bigl[abc (a^2 (b^2 - 1)c + a(- b^2 (c^2 + 1) + 8 b c + c^2 + 1) + (b^2 - 1) c)\bigr]\,
        \frac{1}{t^2} +\\
        &+ \frac{D}{16}\,
        \bigl[a^2 (b^2 - 1)c^4 + a^2 (b^2 - 1) + ac^3 (a^2 (b^4 - 4 b^2 + 1) + 6ab(b^2 - 1) + \\
        &+ b^4 - 4 b^2 + 1) + ac (a^2 (b^4 - 4 b^2 + 1) + 6ab(b^2 - 1) + b^4 - 4 b^2 + 1) + \\
        &- c^2 (6 ab^3 (a^2 + 1) - 6 ab (a^2 + 1)  + 4 a^2 + (a^4 + 4 a^2 + 1) b^4 - (a^4 - 24 a^2 + \\
        &+ 1) b^2)\bigr]\,
        \frac{1}{t}+ \order{1}\,,
    \end{aligned}
    \nonumber\\
    &\mathcal{I}_6 = 
    -D \,a^2 b^2 c^2\,\frac{1}{t^2}+\order{1}\,.
\end{align}
Summing all these contributions, with their multiplicity (see Section \ref{conisec}), we get the full expansion of the SCI for small $t$ and this one no longer suffers of the previous divergences:
\begin{equation}
    \begin{aligned}
    \mathcal{I}_{T^{1,1}} &= 1 + 
    \biggl[\biggl(a^2 + \frac{1}{a^2} + 1\biggr) b^2 + \frac{(a^2 + 1) (c^2 + 1)}{a c} + \frac{c^2 + \frac{1}{c^2} + 1}{b^2}\biggr] \, t^2 + \\
    &+ \biggl[a^4 b^4 + \frac{b^4}{a^4} + \frac{a^3 (c^2 + 1) b^2}{c} + \frac{(c^2 + 1) b^2}{a^3 c} + a^2 \left( (b^2 + 1)^2 + 2 c^2 + \frac{2}{c^2}\right) + \\
    &+ \frac{(b^2 + 1)^2 + 2 c^2 + \frac{2}{c^2}}{a^2} + \frac{a (c^2 + 1) (c^2 b^4 + c^4 + 1)}{b^2 c^3} + \frac{(c^2 + 1) (c^2 b^4 + c^4 + 1)}{a b^2 c^3} + \\
    &+\frac{c^8 + (b^2 + 1)^2 c^6 + 2 (b^8 + b^4 + 1) c^4 + (b^2 + 1)^2 c^2 + 1}{b^4 c^4}\biggr]\, t^4 + \\
    &+ \frac{1}{(abc)^6} \biggl[
    a^{12} c^6 b^{12} + c^6 b^{12} + a^{11} c^5 (c^2 + 1) b^{10} + a c^5 (c^2 + 1) b^{10} + a^{10} c^4 (2 c^4 + \\
    &+ (b^2 + 1)^2 c^2 + 2) b^8 + a^2 c^4 (2 c^4 + (b^2 + 1)^2 c^2 + 2) b^8 + a^9 c^3 (c^2 + 1) (2 c^4 + \\
    &+ (b^4 + 2 b^2 - 1) c^2 + 2) b^6 + a^3 c^3 (c^2 + 1) ( 2 c^4 + (b^4 + 2 b^2 - 1) c^2 + 2) b^6 + \\
    &+a^8 c^2 (2 c^8 + (b^4 + 4 b^2 + 1) c^6 + (2 (b^6 + b^4 + b^2 + 2) b^2 + 1) c^4 + (b^4 +4 b^2 +\\
    &+ 1) c^2 + 2) b^4 + a^4 c^2 (2 c^8 + (b^4 + 4 b^2 + 1) c^6 + (2 (b^6 + b^4 + b^2 + 2) b^2 + 1) c^4 +\\
    &+ (b^4 + 4 b^2 + 1) c^2 + 2) b^4 + a^7 c (c^2 + 1) (c^8 + b^2 (b^2 + 2) c^6 + (b^8 + 4 b^6 - 3 b^4 +\\
    &+ 2 b^2 + 1) c^4 + b^2 (b^2 + 2) c^2 + 1) b^2 +a^5 c (c^2 + 1) (c^8 + b^2 (b^2 + 2) c^6 + (b^8 + 4 b^6 +\\
    &- 3 b^4 + 2 b^2 + 1) c^4 + b^2 (b^2 + 2) c^2 + 1) b^2 + a^6 (c^{12} + (b^2 + 1)^2 c^{10} + (b^8 + 4 b^6 +\\
    &+ 2 b^4 + 2 b^2 + 2) c^8 + 2 (b^4 + 1) (b^8 + 2 b^6 - 2 b^4 + 2 b^2 + 1) c^6 + \\
    &+ (b^8 + 4 b^6 + 2 b^4 + 2 b^2 + 2) c^4 + (b^2 + 1)^2 c^2 + 1)
    \biggr] t^6 + \order{t^7}\,.
    \end{aligned}
\end{equation}

\subsection*{SPP }
\label{sppapp}

This paragraph is divided in two parts.
In the first part we present the explicit results for general fugacities, where we focused only on the “tachyonic” terms.
In the second one we present the explicit results for equal fugacities, where we performed a high order expansion.

In the case of general fugacities we choose the parametrization \eqref{paraspp}.
Combining via \eqref{baf} the contributions for small $t$ of each element of \eqref{sppmiracle}, with their multiplicities (see Section \ref{sppsec}), we obtain an expansion with negative powers of $t$. We list below some of the coefficients $C_{n}$ in front of $t^n$:
\begin{align}
    &
    \begin{aligned}
        C_{-5}&= \frac{(c-d) \left(2 a^2 b^2 c d-a-b\right)}{2 a b}\,,\\
    \end{aligned}\nonumber \\
    &
    \begin{aligned}
        C_{-4}&= \frac{1}{2 a^2 b^2 \left(-a^2 b^2 c d+a+b\right)^2} \bigl(8 a^9 b^9 c^5 d^5 (c-d)-2 a^8 b^7 c^4 d^4 (7 b+11 (c-d)) +\\
        &+2 a^7 b^5 c^3 d^3 (-11 b^3 c d (c-d)+19 b+10 (c-d))+2 a^6 b^3 c^2 d^2 (19 b^4 c d+\\
        &+25 b^3 c d (c-d)-17 b-3 c+3 d)+a^5 b^2 c d (20 b^5 c^2 d^2 (c-d)-78 b^3 c d+\\
        &-41 b^2 c d (c-d)+10)+a^4 b^2 c d (-34 b^4 c d-41 b^3 c d (c-d)+50 b+16 (c-d))+\\
        &+a^3 (6 b^6 c^2 d^2 (d-c)+50 b^4 c d+32 b^3 c d (c-d)-10 b-3 c+3 d)+a^2 b (10 b^4 c d+\\
        &+16 b^3 c d (c-d)-16 b-9 c+9 d)+a b^2 (-10 b-9 c+9 d)+3 b^3 (d-c)\bigr)\,,\\
    \end{aligned}\nonumber \\
    &
    \begin{aligned}
        C_{-3}&=-\frac{1}{2 a^3 b^3 c d (-a^2 c d b^2+b+a)^3}\bigl(
            18 a^{14} c^9 (c-d) d^9 b^{14}-3 a^{13} c^8 (28 b+\\
            &+23 (c-d)) d^8 b^{12}+a^{12} c^5 d^5 (-69 b^3 c^3 (c-d) d^3+352 b c^2 d^2+99 c^2 (c-d) d^2+\\
            &+b^2 (4 c^3-15 d c^2+15 d^2 c-4 d^3)) b^{10}+a^{11} c^4 d^4 (352 c^3 d^3 b^4+234 c^3 (c-d) d^3 b^3+\\
            &-15 (c^3-4 d c^2+4 d^2 c-d^3) b^2-576 c^2 d^2 b+63 c^2 d^2 (d-c)) b^8+a^{10} c^3 d^3 (99 c^4 (c+\\
            &-d) d^4 b^6-15 c d (c^3-4 d c^2+4 d^2 c-d^3) b^5-1280 c^3 d^3 b^4-315 c^3 (c-d) d^3 b^3+\\
            &+(21 c^3-92 d c^2+92 d^2 c-21 d^3) b^2+456 c^2 d^2 b+15 c^2 (c-d) d^2) b^6+\\
            &+a^9 c^2 d^2 (-576 c^4 d^4 b^6-315 c^4 (c-d) d^4 b^5+4 c d (11 c^3-49 d c^2+49 d^2 c-11 d^3) b^4+\\
            &+1812 c^3 d^3 b^3+222 c^3 (c-d) d^3 b^2+(-13 c^3+66 d c^2-66 d^2 c+13 d^3) b+\\
            &-172 c^2 d^2) b^5+6 c (c-d) d b^4+a ((d-c) b^2+36 c d b+24 c (c-d) d) b^3+\\
            &+a^8 c d (-63 c^5 (c-d) d^5 b^8+c^2 d^2 (21 c^3-92 d c^2+92 d^2 c-21 d^3) b^7+1812 c^4 d^4 b^6+\\
            &+438 c^4 (c-d) d^4 b^5-5 c d (9 c^3-47 d c^2+47 d^2 c-9 d^3) b^4-1252 c^3 d^3 b^3-90 c^3 (c+\\
            &-d) d^3 b^2+3 (c^3-7 d c^2+7 d^2 c-d^3) b+24 c^2 d^2) b^3+2 a^2 (c (c-d) d b^5+\\
            &-30 c^2 d^2 b^4-21 c^2 (c\!-\!d) d^2 b^3-2 (c\!-\!d) b^2+46 c d b+18 c (c\!-\!d) d) b^2+a^6 (15 c^5 (c+\\
            &-d) d^5 b^8+c^2 d^2 (-13 c^3+66 d c^2-66 d^2 c+13 d^3) b^7-1252 c^4 d^4 b^6-348 c^4 (c+\\
            &-d) d^4 b^5+6 c d (5 c^3-34 d c^2+34 d^2 c-5 d^3) b^4+1272 c^3 d^3 b^3+162 c^3 (c-d) d^3 b^2+\\
            &+(-2 c^3+25 d c^2-25 d^2 c+2 d^3) b-60 c^2 d^2) b^2+a^7 c d (222 c^5 d^4 b^7-c^2 d (235 d^2 b^5+\\
            &+123 b^2-428 d b+18 d^2) b-2 (9 b^3 d^3+d)+c^3 (235 d^2 b^6-2196 d^3 b^5+348 d^4 b^4+\\
            &+18 b^3+18 d^2 b)+c (45 d^4 b^6+123 d^2 b^3+2)+3 c^4 (152 d^4 b^8-74 d^5 b^7-15 d b^6+\\
            &-116 d^3 b^4)) b^2+a^3 (24 c^3 d^3 b^7+18 c^3 (c\!-\!d) d^3 b^6-(2 c^3-25 d c^2+25 d^2 c-2 d^3) b^5+\\
            &-348 c^2 d^2 b^4-126 c^2 (c-d) d^2 b^3-6 (c-d) b^2+92 c d b+24 c (c-d) d) b+\\
            &+a^5 (-90 c^5 d^4 b^7+6 d^3 b^3+c^2 d (123 d^2 b^5+65 b^2-348 d b+42 d^2) b+d+\\
            &-3 c^3 (41 d^2 b^6-424 d^3 b^5+96 d^4 b^4+2 b^3+14 d^2 b)-c (18 d^4 b^6+65 d^2 b^3+1)+\\
            &+c^4 (-172 d^4 b^8+90 d^5 b^7+18 d b^6+288 d^3 b^4)) b+a^4 (3 c d (c^3\!-\!7 d c^2\!+\!7 d^2 c\!-\!d^3) b^8+\\
            &+428 c^3 d^3 b^7+162 c^3 (c-d) d^3 b^6+(-6 c^3+65 d c^2-65 d^2 c+6 d^3) b^5-568 c^2 d^2 b^4+\\
            &-126 c^2 (c-d) d^2 b^3-4 (c-d) b^2+36 c d b+6 c (c-d) d)
        \bigr)\,,\\
    \end{aligned}\nonumber \\
\end{align}
The full expressions of $C_{-2}$, $C_{-1}$, $C_0$ are omitted for brevity, and the resulting expansion for the SCI is
\begin{equation}
    \text{\eqref{baf}}=
    \sum_{n=-5}^{0}C_n\,t^{n} + \order{t^{1/2}} = 
    1 + \order{t^{1/2}}\,,
\end{equation}
where the last equality holds only in the region \eqref{sppslice}.

Lastly we performed this computation in the case of equal fugacities:
$q=t^5$ and $y_j=t^2$.
We list below the contributions coming from each solution
\begin{align}
    &
    \begin{aligned}
        \mathcal{I}_1&=
        \frac{1}{8 \,t^4}+\frac{3}{4 \,t^3}+\frac{11}{4 \,t^2}+\frac{31}{4\, t}
        +\frac{155}{8}+\frac{183\, t}{4}+\frac{825 \,t^2}{8}+\frac{883\, t^3}{4}+ 
        \frac{1805 \,t^4}{4}+ \\
        &+\frac{1783\, t^5}{2}+\frac{6867 \,t^6}{4}+\frac{6455\, t^7}{2}+\frac{23703\, t^8}{4}+\frac{42605\, t^9}{4}+\frac{75225\, t^{10}}{4}+\order{t^{11}}\,,
    \end{aligned}\nonumber \\
    &
    \begin{aligned}
        \mathcal{I}_2&=
        \frac{1}{256 \,t^{13/2}}-\frac{1}{128\, t^6}+\frac{1}{32 \,t^{11/2}}-\frac{5}{128\, t^5}+\frac{29}{256 \,t^{9/2}}-\frac{19}{128\, t^4}+\frac{3}{8 \,t^{7/2}}-\frac{69}{128\, t^3}+ \\
        &+\frac{275}{256\, t^{5/2}}-\frac{205}{128\, t^2}+\frac{721}{256 \,t^{3/2}}-\frac{137}{32 \,t}+\frac{1795}{256\, t^{1/2}}-\frac{671}{64}+\frac{265\, t^{1/2}}{16}-\frac{1549\, t}{64}+\\
        &+\frac{4737\, t^{3/2}}{128}-\frac{6813 \,t^2}{128}+\frac{2531\, t^{5/2}}{32}-\frac{14423\, t^3}{128}+\frac{10419 \,t^{7/2}}{64}-\frac{1835 \,t^4}{8}+\frac{5203 \,t^{9/2}}{16}+\\
        &-\frac{28947 \,t^5}{64}+\frac{161709 \,t^{11/2}}{256}-\frac{110975 \,t^6}{128}+\frac{38269 \,t^{13/2}}{32}-\frac{207887 \,t^7}{128}+\frac{283031\, t^{15/2}}{128}+\\
        &-\!\frac{381021\, t^8}{128}\!+\!\frac{256445 \,t^{17/2}}{64}\!-\!\frac{684699\, t^9}{128}\!+\!\frac{1825177 \,t^{19/2}}{256}\!-\!\frac{1207957 \,t^{10}}{128}\!+\!\order{t^{21/2}}\,,
    \end{aligned}\nonumber\\
    &
    \begin{aligned}
        \mathcal{I}_3&=
        -\frac{1}{256 \,t^{13/2}}-\frac{1}{128\, t^6}-\frac{1}{32\, t^{11/2}}-\frac{5}{128\, t^5}-\frac{29}{256\, t^{9/2}}-\frac{19}{128\, t^4}-\frac{3}{8\, t^{7/2}}-\frac{69}{128\, t^3}+\\
        &-\frac{275}{256 \,t^{5/2}}-\frac{205}{128\, t^2}-\frac{721}{256 \,t^{3/2}}-\frac{137}{32\, t}-\frac{1795}{256 \,t^{1/2}}-\frac{671}{64}-\frac{265\, t^{1/2}}{16}-\frac{1549\, t}{64}+\\
        &-\frac{4737\, t^{3/2}}{128}-\frac{6813\, t^2}{128}-\frac{2531\, t^{5/2}}{32}-\frac{14423\, t^3}{128}-\frac{10419 \,t^{7/2}}{64}-\frac{1835\, t^4}{8}-\frac{5203 \,t^{9/2}}{16}+\\
        &-\frac{28947\, t^5}{64}-\frac{161709\, t^{11/2}}{256}-\frac{110975\, t^6}{128}-\frac{38269 \,t^{13/2}}{32}-\frac{207887 \,t^7}{128}-\frac{283031\, t^{15/2}}{128}+\\
        &-\!\frac{381021 \,t^8}{128}\!-\!\frac{256445 \,t^{17/2}}{64}\!-\!\frac{684699 \,t^9}{128}\!-\!\frac{1825177 \,t^{19/2}}{256}\!-\!\frac{1207957\, t^{10}}{128}\!+\!\order{t^{21/2}}\,,
    \end{aligned}\nonumber\\
    &
    \begin{aligned}
        \mathcal{I}_4&=
        -\frac{1}{256 \,t^{13/2}}-\frac{1}{128 \,t^6}-\frac{1}{64 \,t^{11/2}}-\frac{5}{128 \,t^5}-\frac{17}{256 \,t^{9/2}}-\frac{19}{128 \,t^4}-\frac{17}{64 \,t^{7/2}}-\frac{57}{128 \,t^3}+\\
        &-\frac{211}{256 \,t^{5/2}}-\frac{169}{128 \,t^2}-\frac{569}{256 \,t^{3/2}}-\frac{115}{32 t}-\frac{1423}{256 \,t^{1/2}}-\frac{567}{64}-\frac{857 \,t^{1/2}}{64}-\frac{1299 t}{64}+\\
        &-\frac{3899 \,t^{3/2}}{128}-\frac{5709 \,t^2}{128}-\frac{4201 \,t^{5/2}}{64}-\frac{12103 \,t^3}{128}-\frac{8673 \,t^{7/2}}{64}-\frac{3091 \,t^4}{16}-\frac{8687 \,t^{9/2}}{32}+\\
        &-\frac{24385 \,t^5}{64}-\frac{135437 \,t^{11/2}}{256}-\frac{93491 \,t^6}{128}-\frac{64273 \,t^{13/2}}{64}-\frac{175131 \,t^7}{128}-\frac{237983 \,t^{15/2}}{128}+\\
        &-\!\frac{321289 \,t^8}{128}\!-\!\frac{215755 \,t^{17/2}}{64}\!-\!\frac{577607 \,t^9}{128}\!-\!\frac{1536829 \,t^{19/2}}{256}\!-\!\frac{1019229 \,t^{10}}{128}\!+\!\order{t^{21/2}}\,,
    \end{aligned}\nonumber\\
    &
    \begin{aligned}
        \mathcal{I}_5&=
        \frac{1}{8 \,t^4}+\frac{3}{4 \,t^3}+\frac{5}{2 \,t^2}+\frac{27}{4 t}+\frac{67}{4}+\frac{79 t}{2}+\frac{709 \,t^2}{8}+\frac{377 \,t^3}{2}+\frac{1535 \,t^4}{4}+\frac{1515 \,t^5}{2}+\\
        &+\frac{11661 \,t^6}{8}+\frac{5473 \,t^7}{2}+\frac{20073 \,t^8}{4}+\frac{36059 \,t^9}{4}+\frac{127299 \,t^{10}}{8}+\order{t^{11}}\,,
    \end{aligned}\nonumber\\
    &
    \begin{aligned}
        \mathcal{I}_6&=
        \frac{1}{256 \,t^{13/2}}-\frac{1}{128 \,t^6}+\frac{1}{64 \,t^{11/2}}-\frac{5}{128 \,t^5}+\frac{17}{256 \,t^{9/2}}-\frac{19}{128 \,t^4}+\frac{17}{64 \,t^{7/2}}-\frac{57}{128 \,t^3}+\\
        &+\frac{211}{256 \,t^{5/2}}-\frac{169}{128 \,t^2}+\frac{569}{256 \,t^{3/2}}-\frac{115}{32 t}+\frac{1423}{256 \,t^{1/2}}-\frac{567}{64}+\frac{857 \,t^{1/2}}{64}-\frac{1299 t}{64}+\\
        &+\frac{3899 \,t^{3/2}}{128}-\frac{5709 \,t^2}{128}+\frac{4201 \,t^{5/2}}{64}-\frac{12103 \,t^3}{128}+\frac{8673 \,t^{7/2}}{64}-\frac{3091 \,t^4}{16}+\frac{8687 \,t^{9/2}}{32}+\\
        &-\frac{24385 \,t^5}{64}+\frac{135437 \,t^{11/2}}{256}-\frac{93491 \,t^6}{128}+\frac{64273 \,t^{13/2}}{64}-\frac{175131 \,t^7}{128}+\frac{237983 \,t^{15/2}}{128}+\\
        &-\!\frac{321289 \,t^8}{128}\!+\!\frac{215755 \,t^{17/2}}{64}\!-\!\frac{577607 \,t^9}{128}\!+\!\frac{1536829 \,t^{19/2}}{256}\!-\!\frac{1019229 \,t^{10}}{128}\!+\!\order{t^{21/2}}\,,
    \end{aligned}\nonumber\\
    &
    \begin{aligned}
        \mathcal{I}_7&=
        -\frac{1}{256 \,t^{13/2}}+\frac{1}{128 \,t^6}\!-\!\frac{3}{128 \,t^{11/2}}+\frac{5}{128 \,t^5}-\frac{21}{256 \,t^{9/2}}+\frac{11}{128 \,t^4}\!-\!\frac{21}{128 \,t^{7/2}}+\!\frac{15}{128 \,t^3}\!+\\
        &-\frac{65}{256 \,t^{5/2}}+\frac{11}{128 \,t^2}-\frac{121}{256 \,t^{3/2}}+\frac{1}{16 t}-\frac{263}{256 \,t^{1/2}}+\frac{1}{32}-\frac{135 \,t^{1/2}}{64}-\frac{7 t}{16}+\\
        &-\frac{123 \,t^{3/2}}{32}-\frac{267 \,t^2}{128}-\frac{879 \,t^{5/2}}{128}-\frac{713 \,t^3}{128}-\frac{835 \,t^{7/2}}{64}-\frac{765 \,t^4}{64}-\frac{407 \,t^{9/2}}{16}+\\
        &-\frac{775 \,t^5}{32}-\frac{12181 \,t^{11/2}}{256}-\frac{6299 \,t^6}{128}-\frac{5477 \,t^{13/2}}{64}-\frac{12443 \,t^7}{128}-\frac{9761 \,t^{15/2}}{64}+\\
        &-\!\frac{23333 \,t^8}{128}\!-\!\frac{34953 \,t^{17/2}}{128}\!-\!\frac{42099 \,t^9}{128}\!-\!\frac{123741 \,t^{19/2}}{256}\!-\!\frac{74759 \,t^{10}}{128}\!+\!\order{t^{21/2}}\,,
    \end{aligned}\nonumber\\
    &
    \begin{aligned}
        \mathcal{I}_8&=
        \frac{1}{8 \,t^2}+\frac{1}{2 t}+\frac{5}{4}+\frac{11 t}{4}+\frac{49 \,t^2}{8}+\frac{55 \,t^3}{4}+\frac{233 \,t^4}{8}+\frac{229 \,t^5}{4}+\frac{435 \,t^6}{4}+\frac{819 \,t^7}{4}+\\
        &+380 \,t^8+\frac{2747 \,t^9}{4}+\frac{4837 \,t^{10}}{4}+\order{t^{11}}\,,
    \end{aligned}\nonumber\\
    &
    \begin{aligned}
        \mathcal{I}_9&=
        \frac{1}{256 \,t^{13/2}}+\frac{1}{128 \,t^6}+\frac{3}{128 \,t^{11/2}}+\frac{5}{128 \,t^5}+\frac{21}{256 \,t^{9/2}}+\frac{11}{128 \,t^4}+\frac{21}{128 \,t^{7/2}}+\frac{15}{128 \,t^3}+\\
        &+\frac{65}{256 \,t^{5/2}}+\frac{11}{128 \,t^2}+\frac{121}{256 \,t^{3/2}}+\frac{1}{16 t}+\frac{263}{256 \,t^{1/2}}+\frac{1}{32}+\frac{135 \,t^{1/2}}{64}-\frac{7 t}{16}+\\
        &+\frac{123 \,t^{3/2}}{32}-\frac{267 \,t^2}{128}+\frac{879 \,t^{5/2}}{128}-\frac{713 \,t^3}{128}+\frac{835 \,t^{7/2}}{64}-\frac{765 \,t^4}{64}+\frac{407 \,t^{9/2}}{16}+\\
        &-\frac{775 \,t^5}{32}+\frac{12181 \,t^{11/2}}{256}-\frac{6299 \,t^6}{128}+\frac{5477 \,t^{13/2}}{64}-\frac{12443 \,t^7}{128}+\frac{9761 \,t^{15/2}}{64}+\\
        &-\frac{23333 \,t^8}{128}+\frac{34953 \,t^{17/2}}{128}-\frac{42099 \,t^9}{128}+\frac{123741 \,t^{19/2}}{256}-\frac{74759 \,t^{10}}{128}+\order{t^{21/2}}\,.
    \end{aligned}
\end{align}
Summing all these contributions, with their multiplicity, we get the full expansion of the SCI for small $t$ and this one no longer suffers of the previous divergences:
\begin{equation}
    \mathcal{I}_{\mathrm{SPP}} = 
    16 \left[ \,
        \sum_{\mu=1}^{6} \mathcal{I}_\mu +
        2\sum_{\mu=7}^{9} \mathcal{I}_\mu \,
        \right] =
        1 + 5 t^4 + 4 t^6 + 20 t^8 + 6 t^9 + 18 t^{10} + \order{t^{11}}\,.
\end{equation}

\bibliographystyle{JHEP}
\bibliography{ref.bib}
\end{document}